\documentclass[a4paper,fleqn,usenatbib]{mnras}

\usepackage[T1]{fontenc}
\usepackage{ae,aecompl}

\pdfoutput=1

\usepackage{savesym}
\usepackage{graphicx}	
\usepackage{amsmath}	
\savesymbol{iint}
\usepackage{amssymb}
\usepackage{txfonts}
\restoresymbol{TXF}{iint}
\usepackage[pdftex]{}
\usepackage{multicol}        
\usepackage{bm}	
\title{The dust effects on galaxy scaling relations}
\author[B. A. Pastrav]{Bogdan A. Pastrav$^{1}$\thanks{E-mail:bapastrav@spacescience.ro}
\\
$^{1}$Cosmology and Astroparticle Physics Laboratory, Institute of Space Science, Atomistilor 409, 077125, Bucharest-Magurele, Romania\\
}
\date{Accepted 2020 February 12. Received 2019 December 13; in original form 2019 June 27}

\pubyear{2020}

\begin{document}
\label{firstpage}
\pagerange{\pageref{firstpage}--\pageref{lastpage}}
\maketitle

\begin{abstract}
Accurate galaxy scaling relations are essential for a successful model of galaxy formation and evolution
as they provide direct information about the physical mechanisms of galaxy assembly over cosmic time.
We present here a detailed analysis of a sample of nearby spiral galaxies taken from the KINGFISH survey. 
The photometric parameters of the morphological components are obtained from bulge-disk decompositions using 
GALFIT data analysis algorithm, with surface photometry of the sample done beforehand. Dust opacities are determined using a 
previously discovered correlation between the central face-on dust opacity of the disk and the stellar
mass surface density. The method and the library of numerical results previously obtained are used to correct the 
measured photometric and structural parameteres for projection (inclination), dust and decomposition effects in order 
to derive their intrinsic values.
Galaxy disk scaling relations are then presented, both the measured (observed) and the intrinsic (corrected) ones, in the 
optical regime, to show the scale of the biases introduced by the aforementioned effects. The slopes of the size-luminosity 
relations and the dust vs stellar mass are in agreement with values found in other works. We derive mean dust optical depth 
and dust/stellar mass ratios of the sample, which we find to be consistent with previous studies of nearby spiral galaxies.
While our sample is rather small, it is sufficient to quantify the influence of galaxy environment (dust, in this case) 
when deriving scaling relations.
\end{abstract}

 \begin{keywords}
  galaxies: spiral -- galaxies: photometry -- galaxies: structure -- ISM: dust, extinction -- galaxies: evolution -- galaxies: fundamental parameters
 \end{keywords}

 \maketitle

\section{Introduction}

Galaxy scaling relations - the understanding of their nature and why they exist - are of pivotal importance in a succesful theory/model of galaxy formation
and evolution. Thus, a valid semi-analytic model or a numerical simulation should be able to predict with great accuracy the characteristics of galaxy 
scaling relations, such as slope, zero point and scatter, at any wavelength. These relations are also important because they provide direct insights 
into the physical mechanisms of how galaxies and their main components assemble over cosmic time.
Basic scaling relations (e.g. size-luminosity/surface brighness, Tully-Fisher (Tully \& Fisher 1977), 
Faber-Jackson (Faber \& Jackson 1976), Kormendy (Kormendy 1977) relation, the Fundamental Plane (Djorkovsky \& Davis 1987), star-formation
vs stellar mass, etc.) rely on photometric and structural parameters such as disk scalelength, bulge effective radius and S\'{e}rsic index, 
luminosity / surface brightness, absolute magnitudes, circular velocity / velocity dispersion, bulge-to-disk ratio etc.. 
Most of these parameters suffer from biases introduced by dust and inclination, especially in spiral galaxies, where dust is 
present in copious quantities in the disk (Tuffs et al. 2002, Popescu et al. 2002, Stickel et al. 2004, Vlahakis et al. 2005, 
Driver et al. 2007, Dariush et al. 2011, Rowlands et al. 2012, Bourne et al. 2012, Dale et al. 2012). These effects are stronger
at shorter wavelengths and at higher inclinations, as already shown by \cite{Tuf04}, \cite{Mol06}, \cite{Gad10}, \cite{Pas13a} and \cite{Pas13b}.
Therefore, one should first remove all these biases when analysing galaxy scaling relations from observational studies.

Previous works to derive intrinsic scaling relations, corrected for the effects of dust and inclination are currently lacking. Among them, \cite{Gra08} is noteworthy.
They used the radiative transfer model of \cite{Pop00} and numerical corrections from \cite{Mol06} for the disk brightness and scale-length, to analyse the 
intrinsic (dust corrected) luminosity-size and (surface-brightness)-size relations for discs and bulges. However, the study of \cite{Mol06} was done for pure disks 
only, at low to intermediate inclinations. Other studies quantifying dust effects on disk photometric parameters are the ones done by \cite{Byu94}, \cite{Eva94} and \cite{Cun01}. 
\cite{Gad10} studied the effects of dust attenuation on both bulge and disk structural parameters, through simulations produced with Monte Carlo radiative transfer 
technique and bulge-disk decompositions. \cite{Gro13} used a correlation between dust opacity and stellar mass surface density that they had identified, together with 
the radiation transfer model of \cite{Pop11} to derive scaling relations for specific star formation rate, 
stellar mass and stellar mass surface density. By removing the biases introduced by dust and projection effects, they were able to reduce the scatter in those relations. 
More recently, \cite{Dev17} presented an inclination-independent technique (linear surface brigthness) to measure sizes and concentrations of infrared selected
samples of disk and flattened eliptical galaxies and showed how structural parameters are biased by projection effects.

In this paper we present the results of a detailed study of a sample of nearby unbarred spiral galaxies from the KINFGISH survey (Kennicutt et al. 2011), 
showing dust biases in disk scaling relations of spiral galaxies. For this purpose, we decompose each galaxy into its main components (bulge+disk),
deriving intrinsic parameters involved in the scaling relations. We follow the method of \cite{Pas13a} and \cite{Pas13b} and use their numerical corrections 
for projection (inclination), dust and decomposition effects. The numerical corrections were derived by analysing 
and fitting simulated images of galaxies produced by means of radiative transfer calculations and the model of \cite{Pop11}. We also use the empirical
relation found by \cite{Gro13} to determine the central face-on dust opacity. This is neccessary when applying the corrections for dust effects.\\
The method presented here is suitable for cases when optical data is available. Our study comes to underline the importance of having accurate, dust-free
scaling relations in models and studies of galaxy formation and evolution (the size-luminosity type relations, stellar mass vs size), or the interstellar medium (ISM) 
evolution (relations such as dust mass vs stellar mass, dust-to-stellar ratio as a function of stellar mass or stellar mass surface density).
The size of our sample is sufficient for the purpose of this work. This study is an application of the results obtained in \cite{Pas13a} and
\cite{Pas13b} and the first which takes all the aforementioned effects to obtain truly intrinsic scaling relations.

The paper is organised as follows. In Sect.~\ref{sec:sample} we present the galaxy sample used in this study, while in Sect.~\ref{sec:method} 
we present the method used for deriving the integrated fluxes, background subtraction and the overall fitting procedure. In Sect.~\ref{sec:dust_opacity} 
we present the relations used to derive dust opacities and masses and then, in sect.~\ref{sec:corr}, our method to correct the derived structural
and photometric parameters for inclination, dust and decomposition effects. In Sect.~\ref{sec:results} we present the main results, plots
with galaxy scaling relations, both observed and intrinsic (dust-free) ones, all the numerical results and comment upon them. In Sect.~\ref{sec:discussion} 
there is a short discussion concerning some of the results, while in Sect.~\ref{sec:conclusions} we summarise the results obtained in this study and draw conclusions.\\
Throughout this paper, where necessary, a Hubble constant of $H_{0}=67.8$ km/s/Mpc (Planck Collaboration 2016) was used.

\section{Sample}\label{sec:sample}

Our sample consists of 18 nearby spiral galaxies, included in the SINGS (\textit{Spitzer} Infrared Nearby Galaxies Survey; Kennicutt et al. 2003) 
survey and the KINGFISH project (Key Insights on Nearby Galaxies: a Far-Infrared Survey with \textit{Herschel}; Kennicutt et al. 2011). 
The KINGFISH project is an imaging and spectroscopic survey, consisting of 61 nearby (d<30 Mpc) galaxies, chosen to cover a wide range of 
galaxy properties (morphologies, luminosities, SFR, etc.) and local ISM environments characteristic for the nearby universe.\\
We extracted the optical images from the NASA/IPAC Infrared Science Archive (IRSA) and NASA IPAC Extragalactic Database (NED). The images were
taken with the KPNO (Kitt Peak National Observatory) and CTIO (Cerro Tololo Inter-American Observatory) telescopes (see Kennicutt et al. (2003). 
From this sample, we exclude elliptical, irregular or dwarf galaxies, as these are not appropriate for the purpose and methods used
in this study. We also exclude barred galaxies. This is done because we want to observe
dust-free scaling relations, and, at this point, we cannot properly account for the effects of dust on the photometric and structural parameters of bars. 
Of course, one could do a 2-component decomposition (disk+bulge) instead of a 3-component one (disk+bulge+bar) for the barred galaxies, but this would bias the 
results obtained for the bulge component parameters, producing an overestimation of $B/T$ or bulge S\'{e}rsic index, as Laurikainen et al. (2006) have shown 
(a fraction of the bar surface brightness could be mixed into the bulge one, while another fraction could be embedded in the disk surface brightness).\\
Therefore, after taking into account these considerations, we are left with a sample of 18 unbarred nearby spiral galaxies, in B band.
We have considered for our study the analysis of galaxy images in the optical regime (B band), as dust and inclination effects are stronger at shorter 
wavelengths (Pastrav et al. (2013a), Pastrav et al. (2013b)), and because our method is tailored for cases where optical data are available.

\section{Method}\label{sec:method}

\subsection{Fitting procedure}\label{sec:fitting}

We used GALFIT (version 3.0.2) data analysis algorithm (Peng et al. 2002, Peng et al. 2010) for 
the fitting procedure of the galaxy images in our sample.  
GALFIT uses a non-linear least-squares fitting based on the Levenberg-Marquardt algorithm. Through this, the
goodness of the fit is checked by computing the $\chi^{2}$ between the
the real galaxy image and the model image (created by GALFIT to fit the galaxy image). This is an iterative
process, and the free parameters corresponding to each component are adjusted
after each iteration to minimize the normalised (reduced) value of
$\chi^{2}$ ($\chi^{2}/N_{DOF}$, with $N_{DOF}$ = number of pixels - number of free
parameters, the number of degrees of freedom). \\
To fit the observed images of the unbarred spiral galaxies and perform bulge-disk decomposition we used the exponential (``expdisk'') and the 
S\'{e}rsic (``sersic'') functions available in GALFIT, for the disk and bulge surface brightness profiles, together with the "sky" function
(for an estimation of the background in each image).
The two functions represent the  distribution of an infinitely thin disk, and their mathematical 
description is given by Eqs.~\ref{eq:exp} and \ref{eq:sersic}, below:
\begin{eqnarray}\label{eq:exp}
\mu(r)=\mu_{0}~exp(-\frac{r}{R_{s,d}})
\end{eqnarray}
\begin{eqnarray}\label{eq:sersic}
\mu(r)=\mu_{0}~exp[-\kappa_{n} (\frac{r}{R_{e,b}})^{1/n}]
\end{eqnarray}
where $\mu_{0}$ is the central surface brightness of the infinitely thin 
disk, $R_{s,d}$ and $R_{e,b}$ are the scale-length of the disk and the effective 
radius of the bulge, $n$ is the S\'{e}rsic 
index, while $\kappa_{n}$ is a variable coupled with $n$ (see Eq.~\ref{eq:kappa_n} or 
Ciotti \& Bertin 1999 and Graham \& Driver 2005). 
\begin{eqnarray}
\kappa_{n}=2n-\frac{1}{3}+\frac{4}{405n}+\frac{46}{25515n^{2}}+O(n^{-3})\label{eq:kappa_n}
\end{eqnarray}
The free parameters of the fits are: the X and Y coordinates of the
centre of the galaxy in pixels, the integrated magnitudes of the disk and bulge components, the
scale-length / effective radius (for exponential/S\'{e}rsic function), axis-ratios, and bulge S\'{e}rsic index (for
S\'{e}rsic function), the sky background (in the preliminary fit - Step 1) and the sky gradients in X and Y. Although the central
coordinates are free parameters, we imposed a constraint on the fitting procedure, ensuring that the bulge and disk components 
were centred on the same position. The axis-ratio is defined as the ratio between the semi-minor and semi-major axis of
the model fit (for each component). The position angle is the angle between the semi-major axis and the Y axis (increasing 
counter clock-wise).

We did not use a ``sigma'' image internally created by GALFIT. Instead, we used a complex star-masking routine 
to eliminate the additional light coming from neighboring galaxies, stars, compact sources, AGN or image artifacts, for each galaxy image, and therefore mask the
corresponding bad pixels. These images were introduced as bad pixel masks in each run of GALFIT.
To create separate images for the components of each galaxy (disk and bulge), we used the functionalities of GALFIT. The images were
needed as a way to determine the value of $B/D$ and compare it with the corresponding value derived from the curve-of-growth (CoG) 
analysis, but also to analyse the fidelity of decomposition. Maps with relative residuals were also created for the same purpose, 
of checking the decomposition.

\subsection{Sky determination and subtraction. Photometry}\label{sec:sky}

To derive the quantities necessary for the galaxy scaling relation analysis we had to calculate the integrated fluxes of 
the galaxies and their constituents. This requires accurate sky subtraction, because systematic errors in the derived sky background 
can propagate into significant uncertainties in the measured structural parameters. This can then bias the bulge-disk decomposition process
towards unphysical and inaccurate results. This was realised in three successive steps.

\textbf{Step 1} We started the fitting process with the sky value as a free parameter, with an initial value obtained from
the image outskirts. The exponential and S\'{e}rsic function parameters were left free as well. The input values for the
coordinates of galaxy centre were determined after a careful inspection of each image. Position angles (PA) and axis-ratios were taken from NED (NASA/
IPAC Extragalactic Database).\\
Then, we ran GALFIT on the full rectangular area of the input image, masked for stars. We check whether the sky background value found by GALFIT is reasonable.
This is necessary as experience shows that the Levenberg-Marquardt
algorithm used by GALFIT can converge on a local minimum, giving inaccurate
background values, when checked against the surface brightness profiles.
This was done from a plot of the average surface brightness of the galaxy, calculated in elliptical annuli (the width of each annulus being 2 arcsecs) 
versus the semi-major axis radius. The profiles flatten to a non-zero level towards the edge of the image, beyond the radius at
which there is no galaxy emission ($R_{max}$). The mean value from this point is our estimate of the sky background. We noticed
that the sky value found by GALFIT in this first run was not always corresponding to the zero galaxy emission level that one would expect.
Therefore we subsequently used our determined sky value for those cases.
We then plotted the sky-subtracted average surface brightness profile, superimposing the model fit and its components (bulge and disk), and the
radius from which the elliptical annuli are incomplete ($R_{2\pi}$).
Subsequently, we calculate and plot the curve-of-growth (CoG) along the semi-major axis-radius, with the background found in GALFIT subtracted, 
and overlay the CoG for disk and bulge. We determine a preliminary value for the bulge-to-disk ratio ($B/D$) from the disk and bulge CoGs.
The $B/D$ value is checked against the one determined by the ratio of the total counts of the decomposed disk and bulge images, because it should 
be consistent, within errors. From the corresponding CoGs, we determine preliminary values for the integrated fluxes of the galaxy, 
the disk and the bulge.  The integrated flux of the galaxy is calculated from the maximum CoG value (in counts), at $R_{max}$ radius
(as stated above, this is the radius beyond which there is no galaxy
emission and, therefore, the CoG is essentially flat toward larger radii). We use the exposure time and flux units conversion (PHOTFLAM) parameters present in the header of each 
galaxy image to convert the fluxes in Jy units.  We then correct the fluxes for foreground extinction, to determine the intrinsic values.

\textbf{Step 2} In the second stage, we do a second run of GALFIT, this time with the sky background fixed to the value found from radial surface brightness profile 
inspection (\textbf{Step 1}). In most cases, it differs slightly from the value found by GALFIT, but it is more accurate and we noticed that these small differences can 
determine significant changes in the values of the output photometric parameters. As before, we plot the average surface brightness profile (calculated in elliptical annuli)
of the observed galaxies, the model and the components versus semi-major axis radius, and the corresponding curves of growth (CoGs). We analyse the
profiles and the fits, and then we determine again the $B/D$ and the new values for the integrated fluxes of the galaxy, the disk and the bulge.

\begin{figure*}
 \begin{center}
  \includegraphics[angle=90,origin=c]{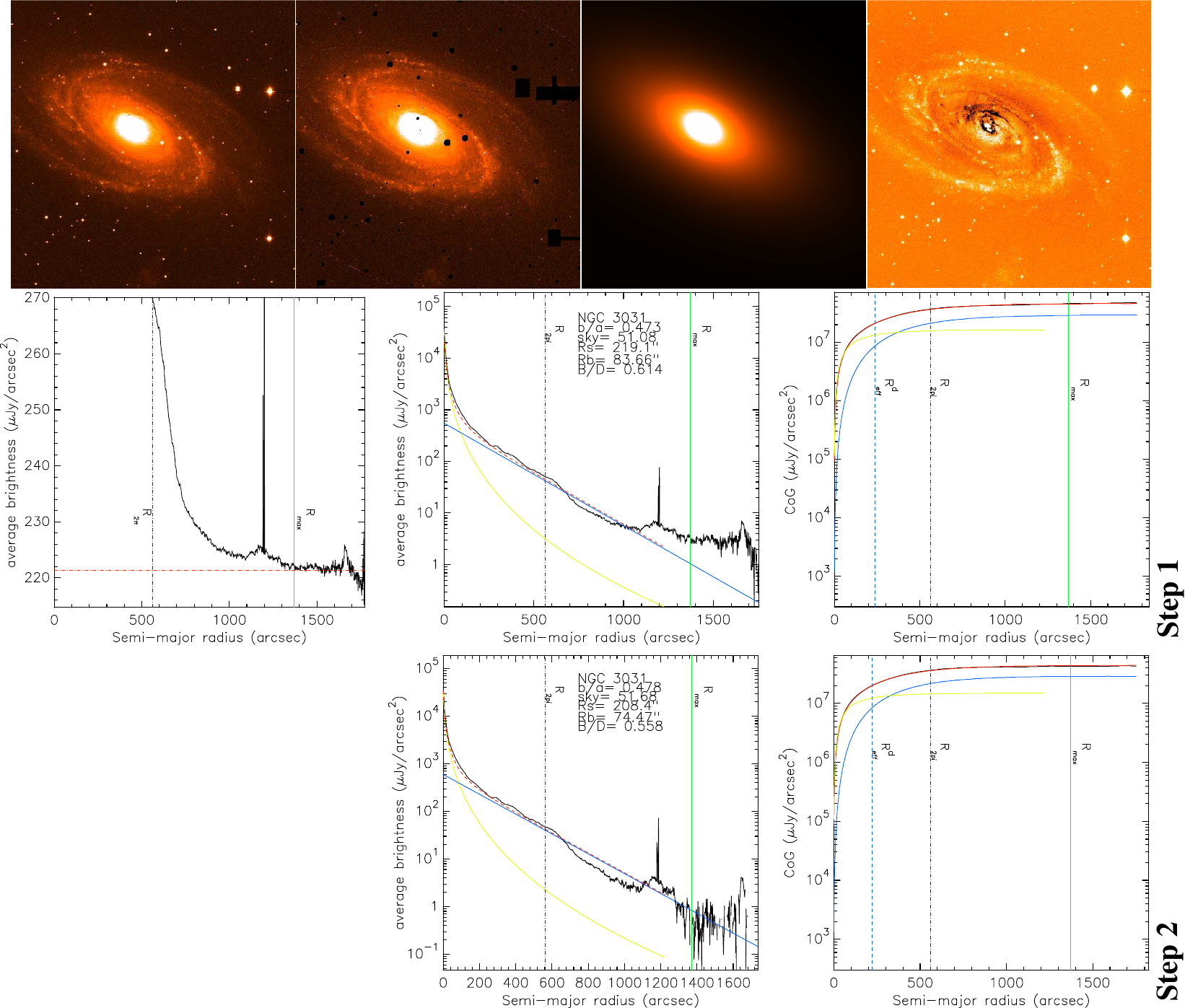}
  \caption{\label{fig:NGC3031} \textbf{NGC3031} The panels are as follows: \textit{Upper row} - the B band observed image, the masked observed image, the best model
  fit image and the absolute residuals image (e.g. data-fit), where very light colored regions/pixels represent positive residuals while dark ones correspond to negative ones; 
  \textit{Middle row} zoom of surface brightness profile in elliptical annuli - used to visually identify the background level
  (drawn as the red dot-dashed line), with $R_{max}$ (radius of the maximum extent of emission in the galaxy) and $R_{2\pi}$ (which denotes the major axis 
  radius out to which data is available over the full azimuthal range) overplotted as vertical dashed dotted violet and green solid lines; surface brightness profiles 
  and CoGs for the observed image (black), model image (the fit - red), disk (blue) and bulge (yellow), corresponding to Step 1, described in Sec.~\ref{sec:sky};
  \textit{Lower row} surface brightness profiles and CoGs corresponding to Step 2. The angular size of the observed image on the sky is $20.22^{\prime}\times20.54^{\prime}$.
  The vertical blue dashed line in the CoG plots show the position of the effective radius of the galaxy. Step 3 fit was not neccesary for this galaxy.}
 \end{center}
\end{figure*}

\begin{figure*}
 \begin{center}
  \includegraphics[angle=90,origin=c]{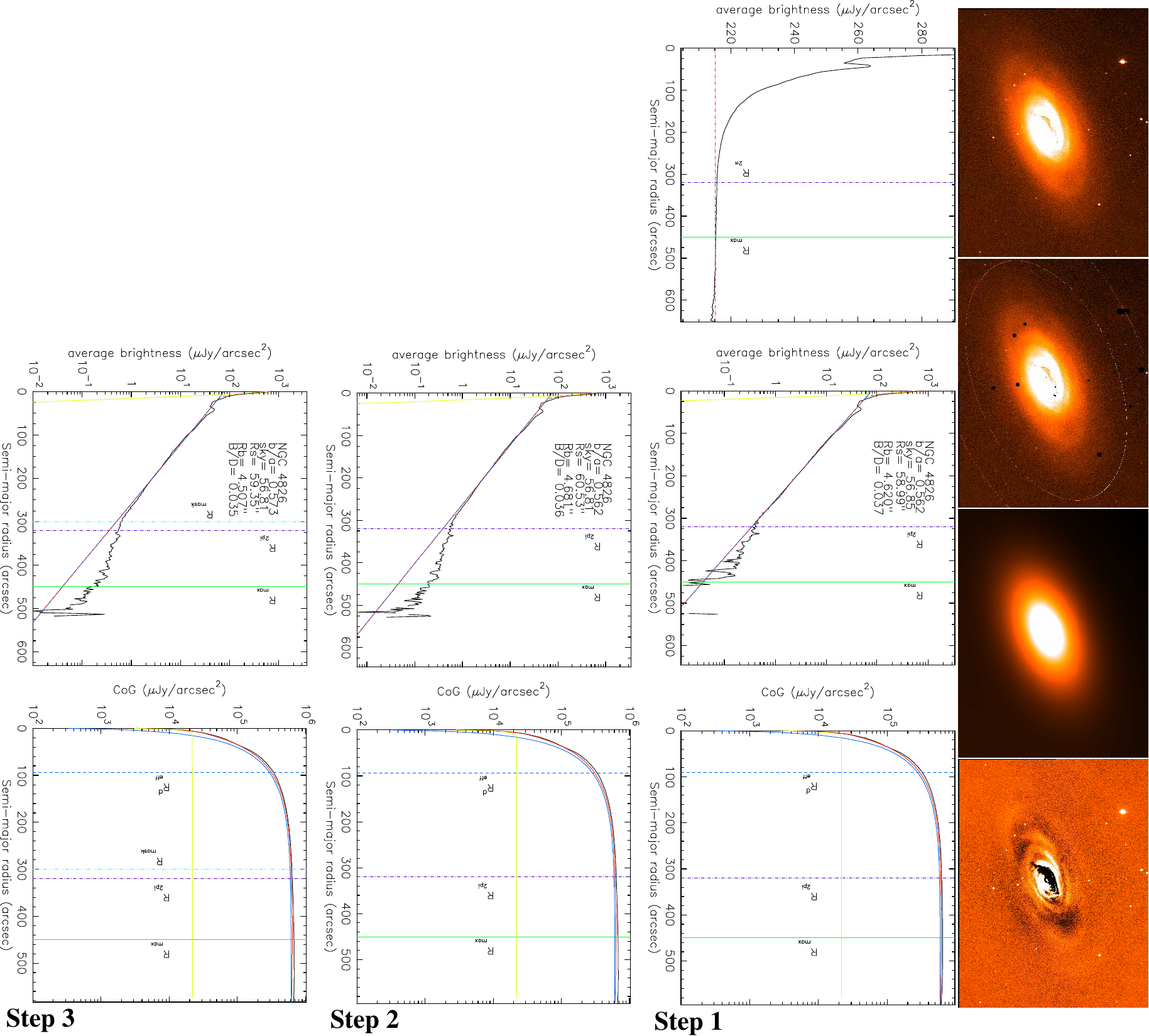}
  \caption{\label{fig:NGC4826} \textbf{NGC4826} As for Fig.~\ref{fig:NGC3031}. Additionally, here the plots corresponding to Step 3 are shown \textit{on the 4th row} - the surface brightness
  profiles and CoGs when an additional mask, setting the weight of all pixels beyond a semi-major radius of $R_{mask}$ to zero is added (light blue vertical line). In the top row - 2nd panel from the left,
  the two dashed white elipses denote $R_{2\pi}$ (inner ellipse, the major axis radius out to which data is available over the full azimuthal range and $R_{max}$ (outer ellipse, 
   the semi-major radius out to which emission from the galaxy could be detected). The angular size of the observed image on the sky is $10.49^{\prime}\times8.03^{\prime}$. }
 \end{center}
\end{figure*}

\textbf{Step 3} (where necessary) If the observed surface brightness profiles showed deviations from an exponential, due to noise or artifacts in the outskirts of the
profiles, we create another mask. The new mask, an elliptical one, contained masked pixels beyond a certain radius ($R_{mask}$) and the masked
pixels from the original. It was used in a third run of GALFIT, with the background fixed to the 2nd run (Step 2) value and all the other parameters free. The whole
process of plotting the surface brightness profiles, CoGs, and calculations of $B/D$ and integrated fluxes is repeated as before.
The uncertainties in the fluxes are estimated from the root mean square of the CoG values from the first 10 elliptical anulli beyond $R_{max}$. 
As in previous steps, the integrated flux of the galaxy is determined from the the CoG value at $R_{max}$, and corrected for foreground extinction.
Finally, after a careful inspection of all the profiles, CoGs, relative residuals, model and decomposed images, together with a check of the $\chi^{2}$ values and the 
structural parameters, we decided which case (Step 1, 2 or 3) is the best fit to each observed galaxy image. Thus, the photometric and structural parameters for that 
case only (either Step 1, 2 or 3 fit) were retained and used further on in our study. For 7 sampled galaxies, we achieved a good fit at Step 1; for 4 of the
others, a better result was found at Step 2; and for the remaining 7 galaxies, we found the best fit at Step 3.
The derived integrated fluxes and bulge-to-disk ratios are given in Table~\ref{tab:photo_fluxes}, for the whole sample, 
together with distances to each galaxy used in this study, 
taken from NASA Extragalactic Database (NED).\\ 
We should mention here the very high derived $B/D$ value for NGC4594 galaxy. This is an edge-on SAa type 
galaxy, with a huge bulge divided by the stellar and dust disks. Thus, the fitting procedure was more complicated, but both the observed and 
intrinsic values of the bulge-to-disk ratio are reasonable.\\
In Figs.~\ref{fig:NGC3031} and \ref{fig:NGC4826} we show examples of the fitting steps for galaxies NGC3031 and NGC4826 to illustrate the whole procedure described here.
For NGC3031, the observed average surface brightness profile was smooth up to $R_{max}$, without deviations from an exponential profile. Therefore,
for this galaxy (as was the case for a few others in our sample), Step 3 fit was not necessary and the corresponding plots are not shown. However, for NGC4826
the situation was different, as it can be seen from the average surface brightness profile (black line), that there are deviations from an exponential
disk profile inside $R_{max}$. Thus, a new mask was used, with all the pixels beyond $R_{mask}$ excluded from the 3rd run fit.\\
We have used the positive sky residuals in the outer parts of galaxies (such as the one noticeable in the surface brightness profiles of
NGC4826 or even NGC3031 towards $R_{max}$ and beyond) to estimate the systematic errors in bulge-to-disk ratios. This is important as the sky level errors dominate the systematic
errors in bulge-to-disk decompositions, as shown by \cite{Sim02}. These are shown in Table~\ref{tab:photo_fluxes}.\\
One could also use a 3rd function during the fitting procedure, to potentially improve the fit in the outskirts of some of the galaxies, as it would be the case
for NGC4826 and a few other galaxies, when various features and deviations from an exponential disk are present, such as truncations / antitruncantions, rings, noise etc. 
However, studying the outskirts of galaxies is beyond the purpose of the present paper, and we judge the 2-component fits to reproduce the surface brightness of the
galaxies with sufficient fidelity in the region of interest for our study (at galactocentric radii up to 5 disk scalelengths).\\

\begin{table}
\caption{\label{tab:photo_fluxes} The calculated fluxes for our sample (B band). The columns represent: (1) - galaxy name; (2) - distance to each galaxy, taken 
from NASA Extragalactic Database (NED), as derived in: $a$ - \protect\cite{Tul13}, $b$ - \protect\cite{Kre17}, $c$ - \protect\cite{Dal09}, $d$ -  
\protect\cite{Jang12}, $e$ - \protect\cite{Man11}, $f$ -  \protect\cite{Poz09}, $g$ -  \protect\cite{Sor14} and $h$ -  \protect\cite{McQ16}; (3) bulge-to-disk 
ratios ($B/D$) derived from the decomposed images, with systematic uncertainties, derived as described in Sec.~\ref{sec:sky}; (4) - the integrated flux for 
each galaxy, in $Jy$; (5) - the error for the galaxy flux; (6), (7) - the integrated fluxes of the disk and bulge components, in $Jy$.}
 \begin{tabular}{{r|r|r|r|r|r|r}}
  \hline \hline
    $Galaxy$   &  $d_{gal}$  &  $B/D$    &   $F_{gal}$  &   $\sigma_{F_{gal}}$  &   $F_{d}$ & $F_{b}$\\
               &   [Mpc]   &                &      [Jy]           &        [Jy]                  &       [Jy]       &      [Jy]     \\
   \hline 
NGC0024  &    $7.67^{a}$  &	  $0.00_{-0.00}^{+0.00}$    &              0.14    &     0.02   &     0.14    &         0.00\\  
NGC0628  &    $9.59^{b}$  &       $0.03_{-0.00}^{+0.00}$    &              1.34      &   0.01   &   1.30      &        0.04\\   
NGC2841  &   $14.60^{a}$  &       $0.19_{-0.03}^{+0.05}$    &        	   0.59       &  0.01   &       0.50  &           0.09\\  
NGC2976  &   $ 3.57^{c}$  &       $0.00_{-0.00}^{+0.00}$    &              0.37      &   0.01   &    0.37     &        0.00\\  
NGC3031  &   $ 3.62^{d}$  &       $0.61_{-0.05}^{+0.06}$    &              5.03      &   0.05   &   3.07      &       1.96\\         
NGC3190  &   $24.20^{e}$  &       $0.36_{-0.08}^{+0.02}$    &              0.25      &   0.01   &     0.18    &         0.07\\  
NGC3621  &   $ 6.73^{a}$  &       $0.05_{-0.01}^{+0.02}$    &              1.51      &   0.08   &     1.45    &         0.06\\  
NGC3938  &   $17.90^{f}$  &       $0.03_{-0.00}^{+0.00}$    &              0.31     &    0.01  &      0.30   &          0.01\\  
NGC4254  &   $14.40^{f}$  &       $0.15_{-0.00}^{+0.00}$    &              0.40      &   0.02	 &    0.35     &       0.05\\  
NGC4450  &   $15.20^{g}$  &       $0.47_{-0.08}^{+0.08}$    &              0.31      &   0.04   &     0.22    &         0.09\\  
NGC4594  &   $ 9.55^{h}$  &       $4.14_{-0.05}^{+0.06}$    &              0.94      &   0.06   &     0.18    &         0.76\\  
NGC4736  &   $ 4.59^{a}$  &       $1.00_{-0.02}^{+0.03}$    &              1.29     &    0.02	 &    0.64     &        0.65\\  
NGC4826  &   $ 5.50^{g}$  &       $0.04_{-0.01}^{+0.00}$    &              0.80	    &    0.06	 &    0.77     &        0.03\\  
NGC5033  &   $19.30^{g}$  &	  $0.30_{-0.02}^{+0.01}$    &              0.33    &     0.06   &    0.25     &       0.08\\  
NGC5055  &   $ 8.20^{g}$  &       $0.16_{-0.00}^{+0.00}$    &              0.82      &   0.02   &    0.70     &       0.12\\  
NGC5474  &   $ 6.98^{a}$  &       $0.17_{-0.02}^{+0.03}$    &              0.10     &    0.01   &     0.08    &        0.01\\      
NGC7331  &   $13.90^{a}$  &       $0.27_{-0.02}^{+0.03}$    &              0.55       &  0.01	 &    0.43     &       0.12\\  
NGC7793  &   $ 3.70^{g}$  &       $0.01_{-0.00}^{+0.00}$    &              1.47       &  0.02   &     1.46    &        0.01\\  
\hline
 \end{tabular}
\end{table}

Having obtained the best fit parameters (following the previously described procedure) and the integrated fluxes for the galaxies and their main components,
we are in the position to calculate the central surface brightness (average effective surface brightness) for disks (bulges), together with the apparent and 
absolute magnitudes for both disks and bulges. Thus, following for example \cite{Gra05} or \cite{Gra08}, for the disk central surface brightness we have
\begin{eqnarray}\label{eq:SB0_d}
 \mu_{0,d}=-2.5\log[(\frac{F_{d}}{2\pi(R_{s,d})^2*Q_{d}})/F_{0}]
\end{eqnarray}
where $F_{0}$ is the zero point magnitude flux, used to convert $\mu_{0}^{d}$ from units of $Jy/arcsec^{2}$ to $mag/arcsec^{2}$ (taken from the header of each 
.fits image), $F_{d}$ is the disk flux (in Jy, see Table~\ref{tab:photo_fluxes}), while $R_{s,d}$ and $Q_{d}$ are the observed disk scalelength (in arcsecs) 
and axis ratio - derived using GALFIT.\\
The absolute and apparent disk magnitudes are obtained using the relations
\begin{eqnarray}\label{eq:Mabs_d}
 M_{d}=m_{d}-25-5\log(d_{gal}/Mpc)
\end{eqnarray}
\begin{eqnarray}\label{eq:mapp_d}
 m_{d}=\mu_{0,d}-2.5\log(2\pi(R_{s,d})^2*Q_{d})
\end{eqnarray}

In a similar way we can write the equations for the bulge effective surface brightness, apparent and absolute magnitudes (see Graham \& Driver 2005 or Graham \& Worley 2008):
\begin{eqnarray}\label{eq:SB0_b}
\mu_{e,b}=-2.5\log[\frac{F_{b}}{2\pi(R_{e,b})^2\exp(\kappa_{n})n\kappa_{n}^{-2n}\Gamma(2n)Q_{b}}/F_{0}] 
\end{eqnarray}
where $F_{b}$ is the integrated flux of the bulge, $R_{e,b}$ is the effective radius (in arcsecs), $\Gamma(2n)=2\gamma(2n,\kappa_{n})$, with $\Gamma$ and $\gamma$ the 
complete and incomplete gamma functions (Graham \& Driver 2005)
\begin{eqnarray}\label{eq:Mabs_b}
 M_{b}=m_{b}-25-5\log(d_{gal}/Mpc)
\end{eqnarray}
\begin{eqnarray}\label{eq:mapp_b}
 m_{b}=\mu_{e,b}-2.5\log(2\pi(R_{e,b})^2Q_{b})-2.5\log[\frac{n\exp(\kappa_{n})\Gamma(2n)}{\kappa_{n}^{2n}}] 
\end{eqnarray}
We need to include the terms $Q_{d}$ and $Q_{b}$ - observed disk and bulge axis-ratios - in Eqs.~\ref{eq:SB0_d}\&\ref{eq:mapp_d} (for disks) and in Eqs.~\ref{eq:SB0_b}\&\ref{eq:mapp_b} 
(for bulges) as both disks and bulges are seen in projection, and we need to take this into account and correct for projection effects. Of course, as shown 
in \cite{Pas13a}, these effects are more pronounced for more inclined (close to edge-on) disks, where the disk thickness becomes relevant.

\subsection{Dust opacity and dust mass derivation}\label{sec:dust_opacity}

We are now in the position to derive the dust central optical depth in B band - $\tau_{B}^{f}$ and subsequently the 
dust mass ($M_{dust}$) for each galaxy, having previously derived all the necessary quantities. In order to do this,
we used the correlation between $\tau_{B}^{f}$ and stellar mass surface density ($\mu_{*}$) of nearby spiral galaxies, 
found by Grootes et al. (2013):  
\begin{eqnarray}\label{eq:Grootes}
\log(\tau_{B}^{f})=1.12(\pm0.11)\cdot\log(\mu_{*}/M_{\odot}kpc^{-2})-8.6(\pm0.8)
\end{eqnarray}
The stellar mass surface density (expressed in units of $M_{\odot}/kpc^{2}$) is derived using their Eq.~4, and replacing 
single S\'{e}rsic effective radius with the disk scale-length derived from bulge-disk decomposition, in that equation.
This relation was obtained by analysing a sample of spiral galaxies taken from the Galaxy and Mass Assembly (GAMA) survey. 
Grootes et al. (2013) showed that with the values obtained through this correlation, they could successfully
correct statistical samples of late-type galaxies for dust attenuation effects when only optical photometric data is available.\\
The stellar masses for our sample are not derived here (in fact being the only quantity involved in the galaxy scaling relations that 
we did not derive in this study) but taken from previous studies of KINGFISH/SINGS galaxies, such as those of \cite{Grossi15},
\cite{Remy15}, \cite{Ski11}, \cite{Ken11} and \cite{Noll09}.

Using Eq.~(2) from Grootes et al. (2013) (but see also Eqs.~(A1-A5) from Appendix A of the same paper) shown here below, we 
derive the dust masses of each galaxy in our sample. This relation was calculated considering the dust geometry of the 
Popescu et al. (2011) model, where the diffuse dust in the disk (which mostly determines the optical depth of a spiral galaxy) 
is distributed axisymetrically in two exponential disks.
\begin{eqnarray}\label{eq:Mdust}
\tau_{B}^{f}=K\frac{M_{dust}}{R_{s,d}}
\end{eqnarray}
where $K = 1.0089pc^{2}/kg$ is a constant containing the details of the dust geometry and the spectral emissivity of the 
Weingartner \& Draine (2001) model, while $R_{s,d}$ is the scale-length of the disk, expressed in kpc.

\begin{table*}
\caption{\label{tab:dust} Dust masses and dust opacities, derived using Eqs.~\ref{eq:Grootes} and \ref{eq:Mdust}. 
The different columns represent: (1) - galaxy name; (2) - B band face-on dust optical depth; (3) - stellar mass surface
densities; (4) - corrected stellar mass surface densities; (5) - stellar masses taken from: $a$ - \protect\cite{Noll09}, $b$ - \protect\cite{Remy15}, 
$c$ - \protect\cite{Grossi15}, $d$ - \protect\cite{Zib11}, $e$ - \protect\cite{Ski11};
(6) - dust masses; (7) - corrected dust masses; (8)-(11) - standard deviation for $\tau_{B}^{f}, \mu_{*}, M_{*}$ and $M_{dust}$. 
The errors for the corrected quantities are the same and thus not given here. In square brackets we have the units 
in which these quantities are expressed. All quantities except dust optical depth are given in decimal logarithm unit scale.}
 \begin{tabular}{{r|r|r|r|r|r|r|r|r|r|r}}
  \hline \hline
 $Galaxy$ &  $\tau_{B}^{f}$ &  $log(\mu_{*})$ & $log(\mu_{*}^{i})$ & $log(M_{*})$ & $log(M_{dust})$ & $log(M_{dust}^{i})$ & $\sigma_{\tau_{B}^{f}}$  & $\sigma_{\mu_{*}}$ & $\sigma_{M_{*}}$ & $\sigma_{M_{dust}}$\\
          &                 &  $[M_{\odot}/kpc^{2}]$ & $[M_{\odot}/kpc^{2}]$  &  $[M_{\odot}]$ &  $[M_{\odot}]$  &  $[M_{\odot}]$  &   &  $[M_{\odot}/kpc^{2}]$  & $[M_{\odot}]$  & $[M_{\odot}]$  \\
  \hline 
NGC0024 &  2.59  &  8.05 &   8.76 &   $9.65^{a}$ &   6.76  &  6.50 &   0.48 &   0.07 &   0.07  &  0.09\\    
NGC0628 &  1.64  &  7.87 &   8.46 &  $10.29^{b}$ &   7.38  &  7.25 &   0.26 &   0.06 &   0.06  &  0.07\\    
NGC2841 &  5.89  &  8.37 &   9.24 &  $10.85^{c}$ &   8.00  &  7.58 &   0.92 &   0.06 &   0.06  &  0.07\\    
NGC2976 &  3.01  &  8.11 &   8.75 &   $9.13^{c}$ &   6.25  &  6.05 &   0.60 &   0.08 &   0.07  &  0.10\\    
NGC3031 &  3.80  &  8.59 &   9.28 &  $11.00^{a}$ &   8.18  &  7.50 &   0.40 &   0.04 &   0.04  &  0.05\\    
NGC3190 &  4.71  &  8.28 &   9.02 &  $10.58^{c}$ &   7.72  &  7.43 &   0.73 &   0.06 &   0.06  &  0.07\\   
NGC3621 &  3.35  &  8.15 &   9.13 &  $10.05^{c}$ &   7.18  &  6.64 &   0.53 &   0.06 &   0.06  &  0.07\\    
NGC3938 &  3.06  &  8.11 &   8.89 &  $10.45^{c}$ &   7.57  &  7.25 &   0.48 &   0.06 &   0.06  &  0.07\\    
NGC4254 &  5.92  &  8.37 &   9.25 &  $10.60^{c}$ &   7.75  &  7.32 &   0.92 &   0.06 &   0.06  &  0.07\\    
NGC4450 &  7.81  &  8.48 &   9.25 &  $10.80^{d}$ &   7.97  &  7.64 &   2.02 &   0.10 &   0.10  &  0.11\\    
NGC4594 &  3.80  &  9.23 &  10.09 &  $10.97^{c}$ &   8.23  &  6.66 &   0.61 &   0.06 &   0.06  &  0.07\\    
NGC4736 &  3.80  &  9.21 &   9.83 &  $10.33^{c}$ &   7.58  &  6.28 &   0.66 &   0.07 &   0.06  &  0.09\\    
NGC4826 &  5.56  &  8.34 &   9.42 &   $9.99^{e}$ &   7.14  &  6.51 &   1.74 &   0.12 &   0.12  &  0.14\\    
NGC5033 &  1.18  &  7.74 &   8.27 &  $10.77^{a}$ &   7.85  &  7.77 &   0.33 &   0.11 &   0.11  &  0.12\\    
NGC5055 &  3.46  &  8.16 &   8.98 &  $10.62^{c}$ &   7.75  &  7.38 &   0.54 &   0.06 &   0.06  &  0.07\\    
NGC5474 &  0.62  &  7.49 &   7.96 &   $9.06^{c}$ &   6.11  &  6.09 &   0.08 &   0.05 &   0.05  &  0.08\\   
NGC7331 &  4.96  &  8.30 &   9.07 &  $10.99^{c}$ &   8.13  &  7.82 &   0.77 &   0.06 &   0.06  &  0.07\\    
NGC7793 &  1.75  &  7.89 &   8.62 &   $9.47^{c}$ &   6.57  &  6.29 &   0.28 &   0.06 &   0.06  &  0.08\\    
\hline
 \end{tabular}
\end{table*}

\subsection{Correcting for dust, projection and decomposition effects}\label{sec:corr}

As we have already underlined in the first section of this paper, obtaining intrinsic structural and photometric parameters is essential
when deriving accurate galaxy scaling relations.\\ 
To correct all the parameters involved we used the method developed and presented in Pastrav et al. (2013a,b). More specifically,
we used the whole chain of corrections presented in Eqs.~(4-13) from Pastrav et al. (2013a) and Eqs.~(3-13) from Pastrav et al. (2013b),
together with all the numerical results (given in electronic form as data tables at CDS) to correct the measured parameters for 
projection (inclination), dust and decomposition effects, in order to obtain their dust-free, intrinsic values.\\
Due to the fact that the numerical corrections are a function of wavelength, dust opacity and/or bulge-to-disk ratio, we used the values of $\tau_{B}^{f}$
and $B/D$ already derived individually for each galaxy and did all the needed interpolations to obtain the final values for the structural and photometric 
parameters. Thus, $Q_{d}$, $R_{s,d}$, $\mu_{0,d}$, $m_{app,d}$, $M_{abs,d}$ (for disk), $B/D$, $n$, $R_{e,b}$, $\mu_{e,b}$, 
$m_{app,b}$, $M_{abs,b}$ (for bulge) were fully corrected. Following from Eqs.~\ref{eq:SB0_d}-\ref{eq:Mabs_d}, we calculate the intrinsic disk central surface brightness as:
\begin{eqnarray}\label{eq:SB0_corr}
 \mu_{0,d}^{i}=\mu_{0,d}-corr(\mu_{0,d})-A_{ext}+A_{dim}
\end{eqnarray}
where $A_{ext}$ is the attenuation due to the foreground galactic extinction (taken from NED, as in Schlafly \& Finkbeiner 2011
recalibration of the Schlegel et al. 1998 infrared based dust map), $corr(\mu_{0,d})$ is the total correction (due to dust and projection effects)
for $\mu_{0,d}$, while $A_{dim}=-2.5\log(1+z)^3$ is the attenuation due to cosmological redshift dimming, per unit frequency interval.\\
The absolute and apparent disk magnitudes are corrected using the relations
\begin{eqnarray}\label{eq:Mabs_corr}
 M_{d}^{i}=m_{d}^{i}-25-5\log(d_{gal}/Mpc)
\end{eqnarray}
\begin{eqnarray}\label{eq:mapp_corr}
 m_{d}^{i}=\mu_{0,d}^{i}-2.5\log(2\pi(R_{s,d}^{i})^2*Q_{d}^{i})
\end{eqnarray}
where $m_{d}^{i}, R_{s,d}^{i}$ and $Q_{d}^{i}$ are the intrinsic apparent disk magnitude, the intrinsic disk scalelength (in arcsecs) and axis-ratio. \\
In a similar way we can rewrite Eqs.~\ref{eq:SB0_b}-\ref{eq:mapp_b} to calculate the intrinsic values for $\mu_{e,b}, m_{b}$ and $M_{b}$ and obtain corresponding corrected scaling relations
for bulges. However, those results will be detailed in a forthcoming paper.\\
All the galaxies in our sample are at low redshift and therefore we did not apply K-corrections or evolutionary ones. Correspondingly, the correction due to cosmological redshift dimming is also quite
small, in the range $0.01-0.05$ mag.\\
Moreover, as $\mu_{*}, M_{dust}, M_{dust}/M_{*}$ do depend on the previously mentioned parameters (namely on $R_{s,d}$), these quantities had to be corrected too for the aforementioned
effects, by replacing the intrinsic parameters in the corresponding equations.\\
The values for all the previously mentioned parameters, for the disk, both the observed and the intrinsic ones, are shown in Tables~\ref{tab:dust} and \ref{tab:photo_struct}, for the 
whole sample. We also give here the results for the bulge component - the measured (observed) ones - in Table~\ref{tab:photo_struct_b}.

\begin{table*}
\caption{\label{tab:photo_struct} The photometric and structural parameters of the disks. The columns represent: (1) - galaxy name; (2) - the intrinsic disk axis-ratio, corrected for inclination (projection) effects and dust 
effects; (3), (4) - the observed and intrinsic disk scalelengths; (5) - intrinsic bulge-to-disk ratio; (6-8) - the observed, the intrinsic disk central surface brightness, and the standard deviation;
(9-11) - the observed, the intrinsic apparent disk magnitude, and the standard deviation; (12-14)- the observed, the intrinsic disk absolute magnitude, and the standard deviation.
In square brackets we have the units in which these quantities are expressed.}
\begin{tabular}{{r|r|r|r|r|r|r|r|r|r|r|r|r|r}}
 \hline \hline
 $Galaxy$  & $Q_{d}^{i}$ & $R_{s,d}$ & $R_{s,d}^{i}$ & $(B/D)^{i}$ & $\mu_{0,d}$ & $\mu_{0,d}^{i}$ & $\sigma_{\mu_{0,d}}$ & $m_{d}$ & $m_{d}^{i}$ & $\sigma_{m_{d}}$ & $M_{d}$ & $M_{d}^{i}$ & $\sigma_{M_{d}}$\\
        &    &  [kpc]  &  [kpc] &   & $[\frac{mag.}{arcsec^{2}}]$ & $[\frac{mag.}{arcsec^{2}}]$  &  & [mag.] & [mag.] &  & [mag.] & [mag.] &  \\
   \hline 
NGC0024   &     0.24 & 1.66 &  1.11 &   0.00  &     19.70  & 17.18  &  1.85 &  11.17  &  9.36  &  1.88  & -18.25  &  -20.07  &  1.97  \\
NGC0628   &     0.95 & 3.61 &  3.29 &   0.03  &     20.33  & 19.64  &  0.54 &   8.79  &  8.45  &  0.55  & -21.12  &  -21.46  &  0.58  \\
NGC2841   &     0.51 & 4.15 &  2.54 &   0.20  &     19.97  & 17.23  &  0.79 &   9.85  &  8.19  &  0.80  & -20.97  &  -22.63  &  0.87  \\
NGC2976   &     0.57 & 0.74 &  0.62 &   0.00  &     19.38  & 17.34  &  1.03 &   9.92  &  8.38  &  1.05  & -17.84  &  -19.39  &  1.13  \\
NGC3031   &     0.47 & 3.73 &  2.90 &   0.63  &     20.74  & 18.31  &  0.59 &   7.86  &  6.04  &  0.60  & -19.94  &  -21.76  &  0.69  \\
NGC3190   &     0.39 & 3.35 &  2.40 &   0.40  &     19.17  & 16.38  &  1.74 &  10.88  &  8.84  &  1.77  & -21.04  &  -23.08  &  1.79  \\
NGC3621   &     0.44 & 2.11 &  1.15 &   0.04  &     18.87  & 16.53  &  0.73 &   8.67  &  7.69  &  0.74  & -20.47  &  -21.45  &  0.80  \\
NGC3938   &     0.92 & 3.50 &  2.42 &   0.03  &     20.32  & 19.14  &  1.03 &  10.38  & 10.01  &  1.05  & -20.88  &  -21.26  &  1.18  \\
NGC4254   &     0.78 & 3.10 &  1.89 &   0.16  &     20.18  & 17.92  &  0.95 &  10.22  &  9.03  &  0.97  & -20.58  &  -21.77  &  1.26  \\
NGC4450   &     0.69 & 3.46 &  2.37 &   0.49  &     20.72  & 17.96  &  1.43 &  10.77  &  8.83  &  1.45  & -20.14  &  -22.08  &  1.65  \\
NGC4594   &     0.11 & 1.50 &  1.10 &   4.21  &     18.43  & 13.53  &  2.16 &  10.94  &  7.41  &  2.21  & -18.96  &  -22.49  &  2.24  \\
NGC4736   &     0.72 & 1.35 &  0.71 &   1.05  &     19.31  & 17.72  &  1.03 &   9.69  &  8.53  &  1.05  & -18.62  &  -19.78  &  1.13  \\
NGC4826   &     0.57 & 1.58 &  0.77 &   0.04  &     19.63  & 17.04  &  0.76 &   9.36  &  8.36  &  0.78  & -19.34  &  -20.34  &  0.86  \\
NGC5033   &     0.43 & 7.03 &  7.12 &   0.31  &     21.41  & 20.61  &  0.92 &  10.72  & 10.12  &  0.94  & -20.71  &  -21.31  &  1.14  \\
NGC5055   &     0.49 & 3.56 &  2.63 &   0.15  &     20.75  & 18.73  &  0.58 &   9.46  &  8.39  &  0.60  & -20.11  &  -21.18  &  0.90  \\
NGC5474   &     0.97 & 0.96 &  1.42 &   0.16  &     21.92  & 21.83  &  1.09 &  11.80  & 11.75  &  1.11  & -17.41  &  -17.47  &  1.27  \\
NGC7331   &     0.36 & 5.38 &  3.66 &   0.29  &     20.38  & 17.19  &  0.76 &   9.99  &  7.62  &  0.77  & -20.72  &  -23.09  &  0.85  \\
NGC7793   &     0.64 & 1.48 &  1.06 &   0.01  &     19.74  & 19.02  &  0.57 &   8.67  &  8.63  &  0.59  & -19.18  &  -19.21  &  0.67  \\
\hline
 \end{tabular}
\end{table*}

\begin{table}
\begin{center}
\caption{\label{tab:photo_struct_b} The photometric and structural parameters of the bulges. The columns represent: (1) - galaxy name; (2) - the observed bulge axis-ratio; (3) - the observed effective radii; 
 (4) - (5) -  the derived effective surface brightness and absolute bulge magnitudes; (6) - bulge S\'{e}rsic index.}
 \begin{tabular}{{r|r|r|r|r|r}}
 \hline  \hline 
$Galaxy$ & $Q_{b}$ & $R_{e,b}$ & $\mu_{e,b}$  & $M_{b}$ & $n$\\
    &   &  [kpc]  & $[\frac{mag.}{arcsec^{2}}]$  & [mag.] &  \\
  \hline
NGC0024   &     0.00 &  0.00 &   0.00  &    0.00  &  0.00\\
NGC0628   &     0.92 &  0.45 &  20.19  &  -17.34  &  1.17\\
NGC2841   &     0.66 &  0.69 &  19.06  &  -19.13  &  1.50\\
NGC2976   &     0.00 &  0.00 &   0.00  &    0.00  &  0.00\\
NGC3031   &     0.70 &  1.47 &  20.84  &  -19.45  &  3.20\\
NGC3190   &     0.27 &  1.53 &  18.54  &  -19.96  &  0.56\\
NGC3621   &     0.53 &  1.15 &  21.29  &  -17.01  &  0.15\\
NGC3938   &     0.94 &  0.45 &  20.22  &  -17.19  &  0.86\\
NGC4254   &     0.79 &  1.65 &  21.97  &  -18.46  &  2.02\\
NGC4450   &     0.62 &  2.18 &  21.75  &  -19.33  &  3.63\\
NGC4594   &     0.75 & 21.31 &  25.87  &  -20.53  &  4.76\\
NGC4736   &     0.90 &  0.27 &  17.89  &  -18.66  &  1.60\\
NGC4826   &     0.66 &  0.12 &  18.28  &  -15.82  &  0.78\\
NGC5033   &     0.40 &  1.76 &  20.18  &  -19.46  &  1.34\\
NGC5055   &     0.62 &  1.32 &  21.21  &  -18.19  &  1.10\\
NGC5474   &     0.77 &  1.04 &  23.82  &  -15.51  &  1.72\\
NGC7331   &     0.39 &  1.25 &  19.35  &  -19.24  &  0.71\\
NGC7793   &     0.85 &  0.03 &  17.73  &  -13.76  &  1.33\\
\hline
 \end{tabular}
 \end{center}
\end{table}

\subsection{Estimation of errors}

The output best-fit parameters given by GALFIT suffer from an underestimation of uncertainties, as shown by \cite{Hau07}. To estimate
the errors on the main photometric parameters, we ran a new set of fits, for a few sampled galaxies. We fixed the sky value to the one found by GALFIT
in Step 1 and added $\pm1\sigma$, or $\pm3\sigma$ ($\sigma$ being the uncertainty in the sky level), leaving all other parameters free. 
The systematic errors in the disk scalelengths and bulge effective
radii were within the range 1-10 pixels (1-3 arcsecs). They were less significant for the axis-ratios, up to $0.01$. Also, the random errors which occur in an
exponential fit are $<10\%$ (Maltby et al. 2012).
The error over $d_{gal}$ (measured distance to the galaxy) was taken from NED. Having the flux uncertainties already calculated (Table~\ref{tab:photo_fluxes}) 
we performed propagation of errors in Eqs.~\ref{eq:SB0_d}-\ref{eq:Mabs_d} and \ref{eq:Grootes}-\ref{eq:mapp_corr} to derive the standard deviations ($\sigma$) for all the needed parameters. 
The respective values can be seen in Tables~\ref{tab:dust} and \ref{tab:photo_struct}. 

\begin{figure*}
\begin{center}
 \includegraphics[scale=0.439]{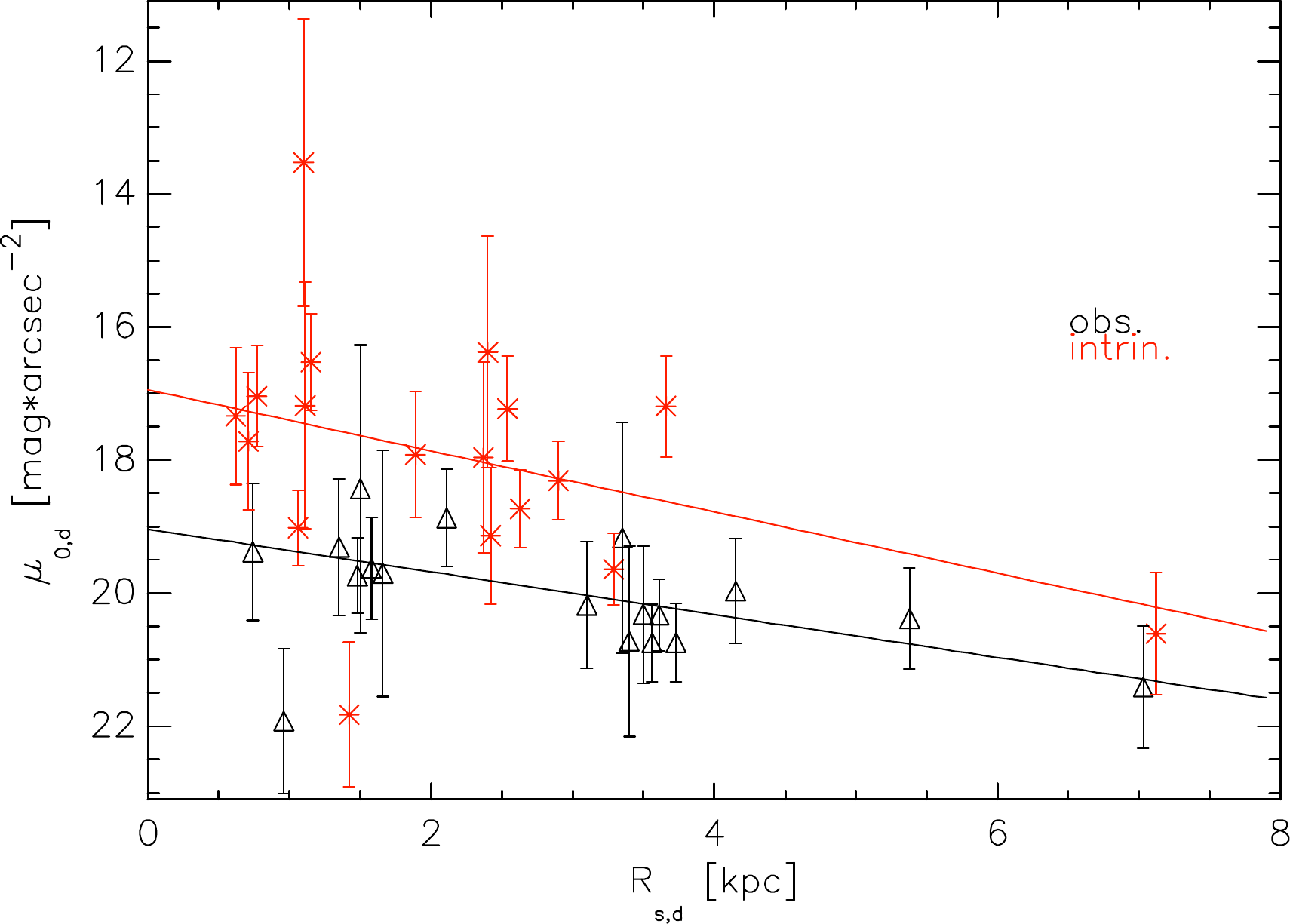}
 \includegraphics[scale=0.44]{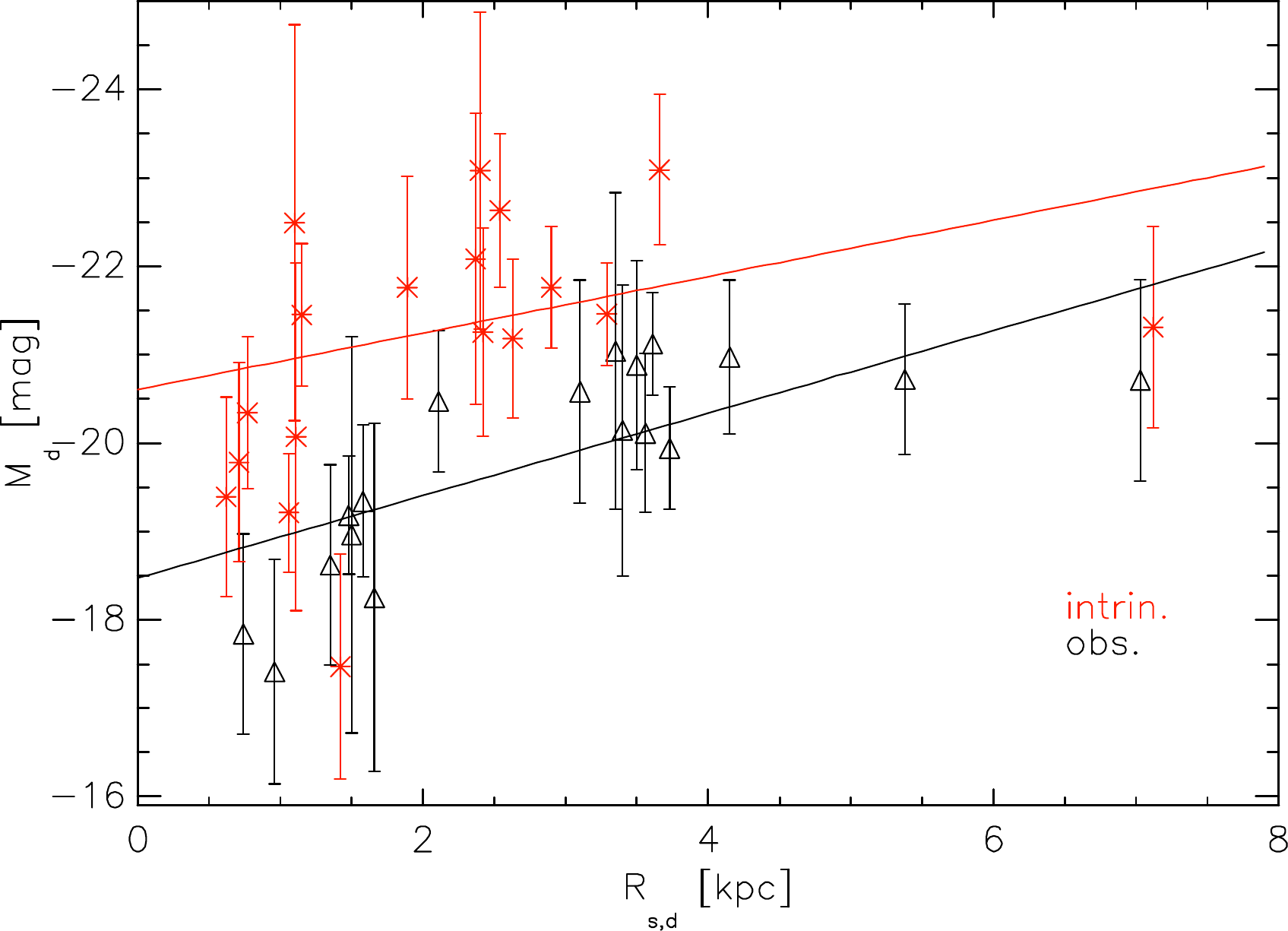}
 \caption{\label{fig:mu0_Mdisk} Left panel: Central disk surface brightness versus disk scalelength. Right panel: Disk magnitude versus disk scalelength. The black triangles represent
 the measured (observed) values, while the red crosses are the intrinsic (corrected for inclination, dust and decomposition effects) values. The black and red solid lines are obtained
 from a linear regression fit to the data points. The error bars represent the standard deviations.}
 \end{center}
\end{figure*}

\section{Results}\label{sec:results}

\begin{figure}
 \includegraphics[scale=0.47]{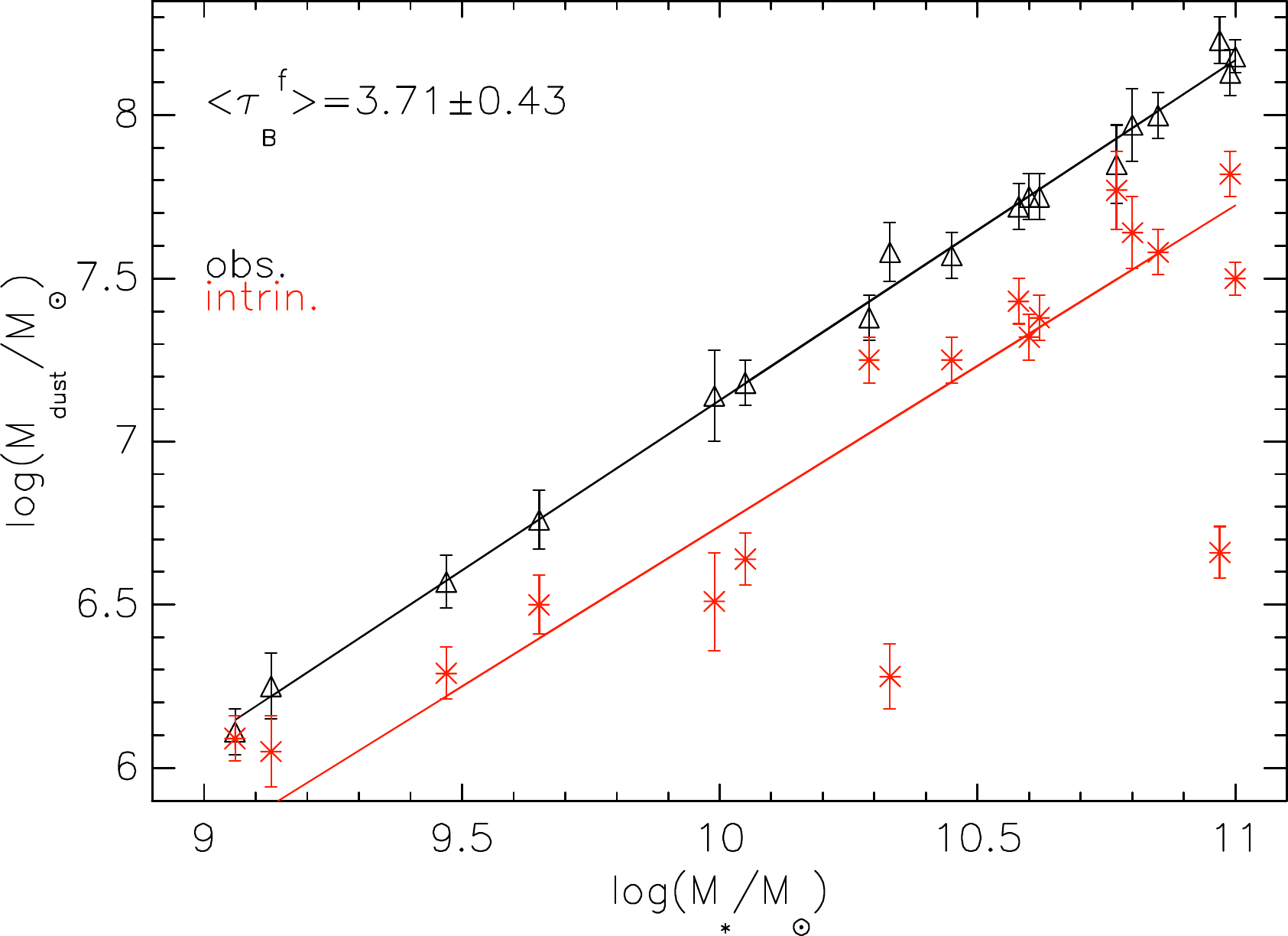}
\caption{\label{fig:Mdust_Mstar} The galaxy dust mass (calculated using Eq.~\ref{eq:Mdust}) versus galaxy stellar mass. The black triangles represent the measured (observed) values, while the red
crosses are the intrinsic (corrected for inclination, dust and decomposition effects) values. The black and red solid lines are obtained from a linear regression fit to the data 
points. The mean value of the B band central optical depth is shown on the plot. The error bars represent the standard deviations.}
\end{figure}

\begin{figure*}
\begin{center}
 \includegraphics[scale=0.47]{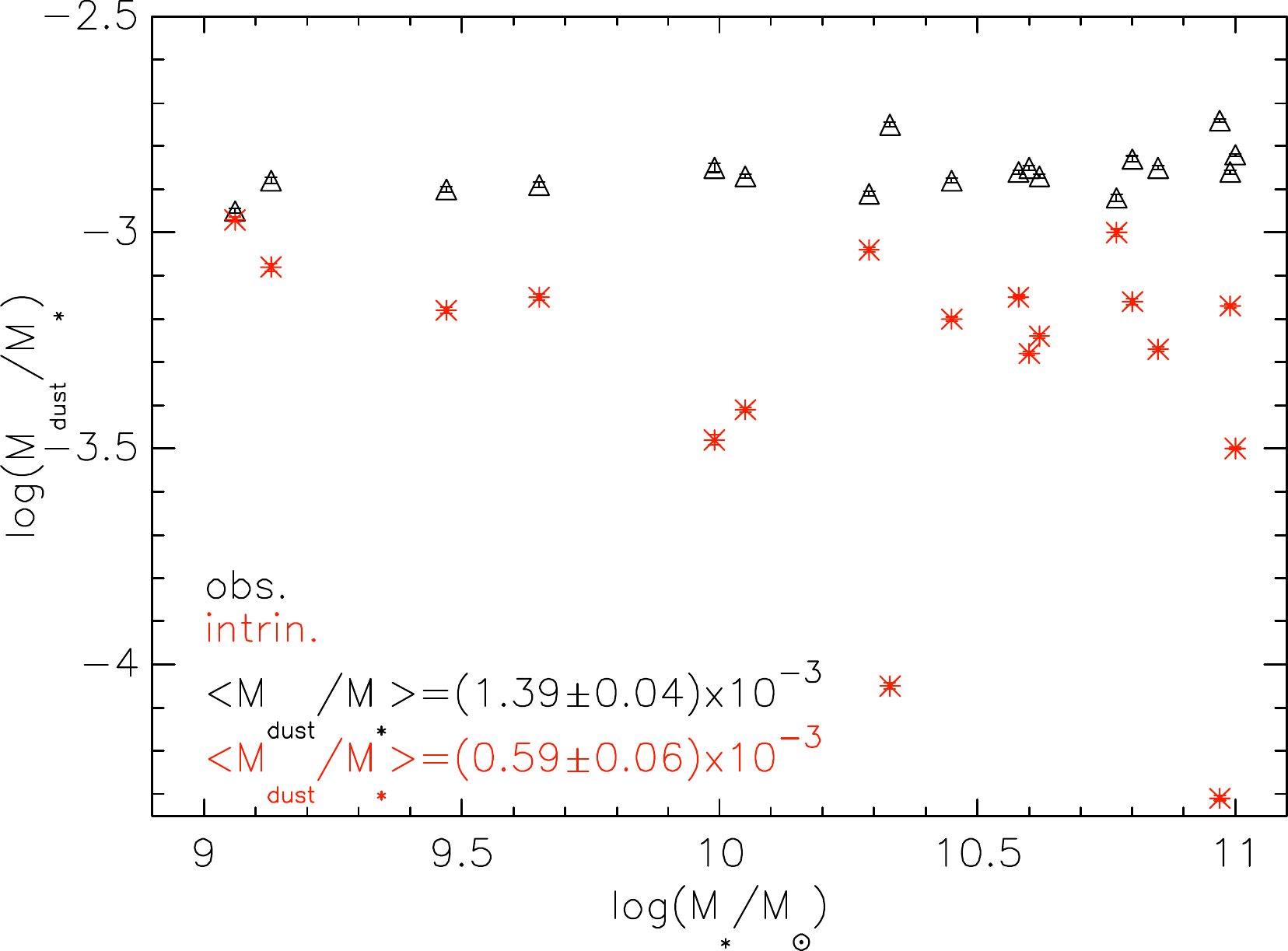}
 \hspace{-0.0cm}
 \includegraphics[scale=0.47]{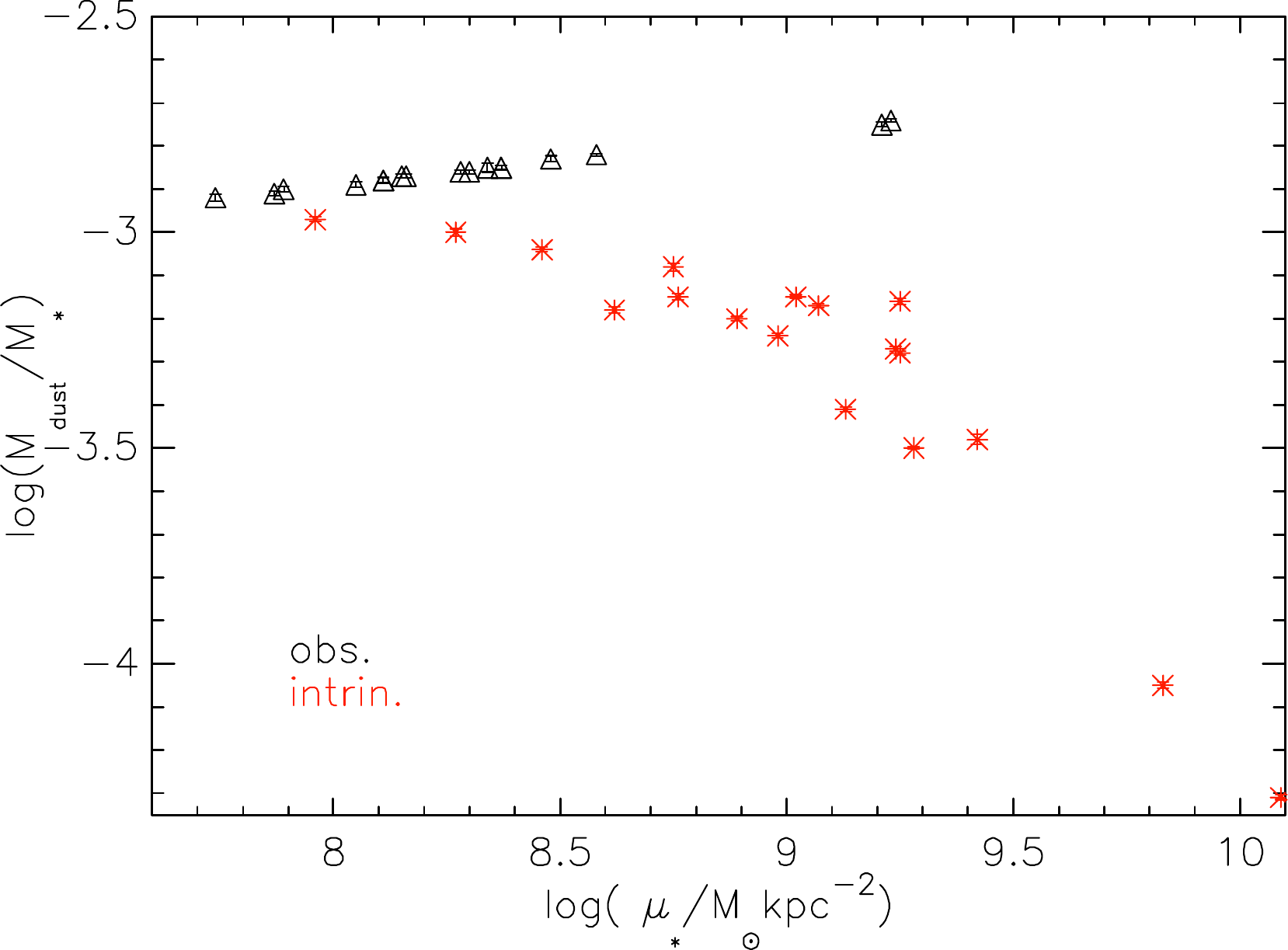}
 \caption{\label{fig:DS_Mstar_ustar} Left panel: Dust to stellar mass ratio, versus galaxy stellar mass. Right panel: Dust to stellar mass ratio versus stellar surface density.
 The black triangles represent the measured (observed) values, while the red crosses are the intrinsic (corrected for inclination, dust and decomposition effects) values. The average
 observed and corrected values for the dust to stellar mass ratio are shown on the left panel plot. The error bars represent the standard deviations.}
 \end{center}
\end{figure*}

\begin{figure}
 \includegraphics[scale=0.47]{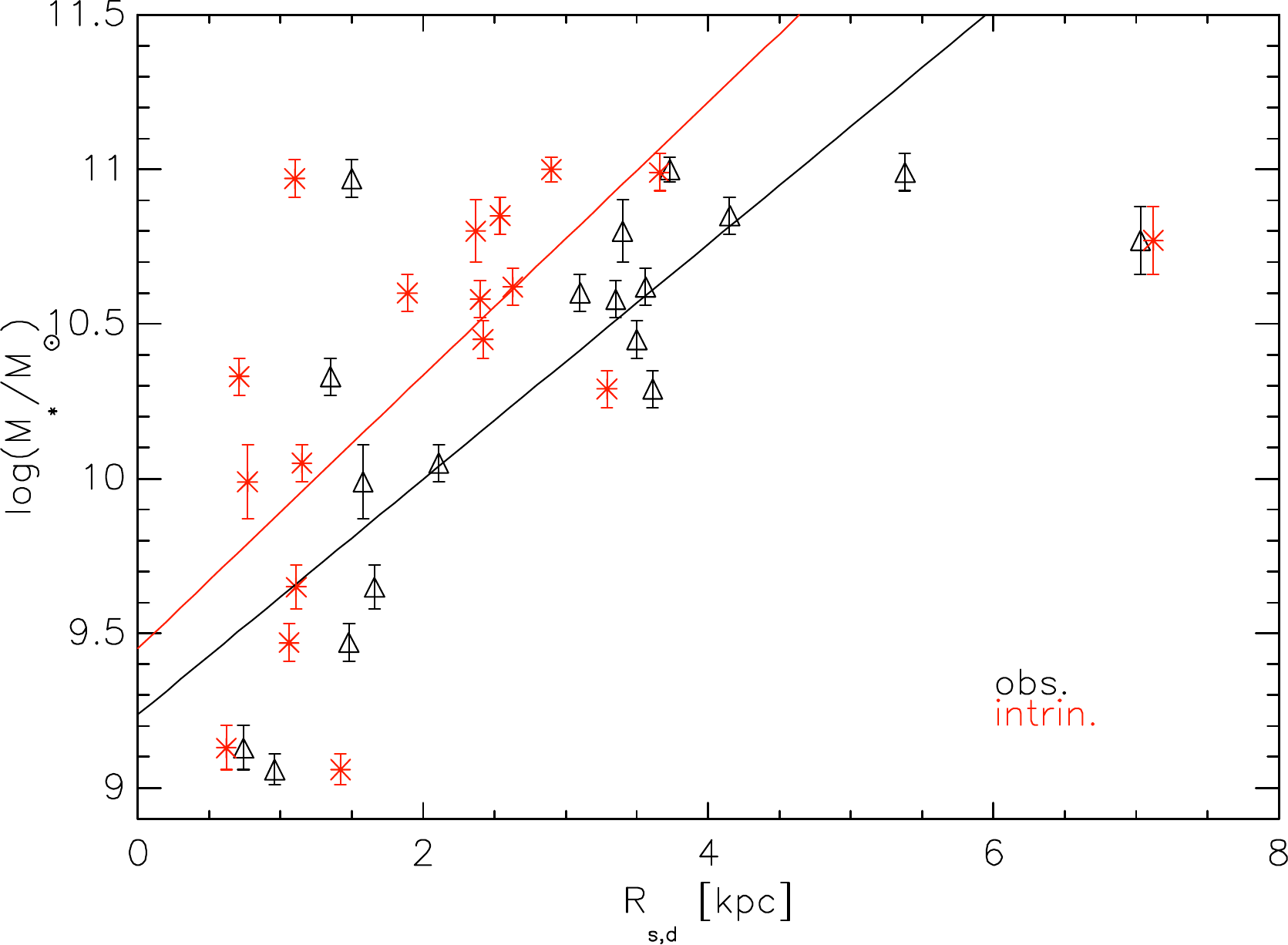}
\caption{\label{fig:Mstar_Rdisk} The galaxy stellar mass as a function of disk scalelength. The symbols and solid lines follow the same convention as in previous plots.}
\end{figure}

In this section, the main results of this study are presented. We concentrate on the scaling relations of the disks. The relations for the bulge component will be 
presented in a forthcoming paper, with scaling relations for elliptical galaxies.
For three galaxies in our sample, namely NGC3031, NGC4594 and NGC4736, the initial value derived for $\tau_{B}^{f}$ is higher than $8.0$ - the upper limit of the numerical corrections for
dust effects given in \cite{Pas13a} and \cite{Pas13b}. These values are unrealistically large. This could be due to either an overestimate of the stellar mass, or issues with the fit/sky
determination (especially for the case of NGC4594) resulting in a larger value for $R_{s,d}$. Therefore, for these 3 galaxies we have chosen to fix the value of the dust
opacity to $3.80$ (see Table \ref{tab:dust}), the average value found by \cite{Dri07}. However, this does not significantly change the overall trends we observe or the mean values
for $\tau_{B}^{f}$ and dust/stellar ratios, even if we exclude these galaxies from calculations.

First, in Fig.~\ref{fig:mu0_Mdisk} we show the brightness-size and size-luminosity relations for disks in our sample, in the form of central surface brightness vs. disk scale-length 
(left panel) and absolute magnitude vs. disk scalelength (right panel). For both cases, we show the observed values (black triangles) and the intrinsic ones (red crosses), fully corrected.
One can notice from the left panel plot a displacement of the corrected values with respect to the measured ones to the left, towards smaller scalelengths, and upwards, towards higher 
central surface brightness values. This is expected because it was shown in \cite{Pas13a} how dust biases the disk measurements towards larger scalelengths and fainter central surface 
brightness values compared with the real (intrinsic) values, with projection and decomposition effects having smaller contributions. 
The same argument explains the similar behaviour of observed vs. intrinsic parameters
seen in the right panel of Fig.~\ref{fig:mu0_Mdisk}. As one can see from both Table~\ref{tab:photo_struct} and Fig.~\ref{fig:mu0_Mdisk} (left panel),
the decrease in central surface brightness is very strong as a result of applying the corrections - up to ~2 mag. or more in some
cases. The corresponding changes in the values of $M_{d}$ are also quite large, but not as significant as for $\mu_{0,d}$. These differences are more significant than in previous similar studies,
such as that of \cite{Gra08} for example.\\
The solid black and red lines represent linear regression fits to the observed and intrinsic data points. Applying the aforementioned corrections produces a 
slope change for both relations displayed in Fig.~\ref{fig:mu0_Mdisk}. Specifically, for the $\mu_{0,d} - R_{s,d}$ relation, the slope changes from $0.32\pm0.07$ to $0.45\pm0.15$, while for
the $M_{d} - R_{s,d}$ relation the slope decreases from $-0.46\pm0.10$ to $-0.32\pm0.14$. The value of $0.32$ is close to the average ones of $0.32$ and $0.35$ found by \cite{Cou07} for 
the size-luminosity relations in I band and K bands, from an extensive analysis of global structural parameters of field and cluster spiral galaxies. 
\cite{Cou07} showed that the slope of size-luminosity relation varies strongly with morphology of the spiral galaxies, with early types having smaller scalelengths than late type spirals. 
Our sample is too small to examine this behaviour.\\

We present in Fig.~\ref{fig:Mdust_Mstar} an important scaling relation for galaxy and ISM evolution studies, the dust mass (calculated from Eq.~\ref{eq:Mdust}) versus the stellar mass relation.
We recover the trend already observed in other studies of disk scaling relations, a linear increase of the dust mass as we go towards galaxies with higher stellar masses. 
Since the dust mass as expressed in Eq.~\ref{eq:Mdust} depends on a parameter which is affected by dust, inclination and decomposition effects - $R_{s,d}$, we also show in red the same relation with the 
corresponding corrections applied. Similar to the first two relations, we find the best fit (solid black and red lines) from a linear regression procedure applied to the observed and corrected
values. The slope of $M_{dust}-M_{*}$ is close to unity for the observed relation - $1.04\pm0.02$, with a small decrease for the corrected one ($0.98\pm0.14$). It is 
also seen that for NGC4594 and NGC4736 the intrinsic dust masses are far from the intrinsic relation and their observed values. This
could be due to our assumption for the dust opacity $\tau_{B}^{f}=3.80$. Overall, an increase in the scatter from the observed to intrinsic relation is noticeable.
This can be explained by the uncertainty involved in deriving an accurate value for the dust optical depth (it is a quantity which in general is difficult to
be determined with great precision) or the underestimation of errors for the stellar mass
(taken from the source papers mentioned in Table~\ref{tab:dust}), dust opacity and disk scalelength.\\
One important result that we should mention here is the B band face-on mean opacity of disks derived for the entire sample (displayed on the plot) from the values in Table~\ref{tab:dust}:
$\tau_{B}^{f}=3.71\pm0.43$. This is a value consistent with other studies of dust attenuation in spiral galaxies.
For example, \cite{Dri07} found a value of $3.8\pm0.7$ by studying the empirical attenuation - inclination relation in B band for a large sample of 10095 galaxies from the
Millennium Galaxy Catalog (MCG), result which agrees well with the theoretical predictions of dust attenuation versus inclination as a function of $\tau_{B}^{f}$ from \cite{Tuf04}.

In connection with the previous plot, in Fig.~\ref{fig:DS_Mstar_ustar} are plotted the dust-to-stellar mass ratio versus stellar mass relation (left panel), while in the right panel the dust-to-stellar
mass ratio is plotted as a function of stellar mass surface density, $\mu_{*}$. As one can see from both plots, the variation of $M_{d}/M_{*}$ with stellar mass or $\mu_{*}$ shows a flat, 
slightly increasing trend. However, after correcting all the parameters involved in these scaling relations (except $M_{*}$) for the previously mentioned effects, we see how the dust-to-stellar
mass ratio anti-correlates with both the stellar mass and stellar mass surface density. These trends are not new and have been previously discovered in larger observational studies by 
\cite{Cort12} and \cite{daC10}, the latter in the form of a $M_{d}/M_{*} - sSFR$ relation ($sSFR=SFR/M_{*}$ - specific star formation rate). \cite{Grossi15} confirmed this behaviour for the dust/stellar
mass ratio as a function of stellar mass from a study of Virgo cluster galaxies and KINGFISH spirals, among others.
As \cite{daC10} explained, the trend seen in Fig.~\ref{fig:DS_Mstar_ustar} is a consequence of the $sSFR$ variation with stellar mass. Thus, for low stellar mass galaxies 
that have high $sSFR$ and gas fractions, a large amount of dust
is formed, exceeding the dust quantity destroyed by various processes in the ISM. When we go towards more massive galaxies (e.g. earlier morphologies, higher $M_{*}$), both the specific star
formation rate and gas fraction decrease, and so the newly formed dust can no longer overcome the destroyed mass of dust. This should explain why $M_{d}/M_{*}$ decreases with stellar mass
and the surface density of the stellar mass.\\
In all the previously mentioned studies, the observed decrease of the dust-to-stellar mass ratio with stellar mass is slightly more pronounced than ours. This is due to the samples being
analysed in the respective studies spanning more orders of magnitude in stellar mass. The fact that we notice the expected trends in Fig.~\ref{fig:DS_Mstar_ustar} only after applying 
the corrections suggest that at least at short wavelengths observational results can be biased and lead to wrong conclusions. Therefore, we caution that when these quantities do depend
on dust biased quantities, the necessary corrections should be applied to the measured quantities.\\
In addition, we derive the mean dust-to-stellar mass ratio for the galaxies in our sample, and display both the observed and corrected values on the left panel plot in Fig.~\ref{fig:DS_Mstar_ustar}.
The measured value of $(1.39\pm0.04)\times10^{-3}$ ($-2.85$ in log scale) agrees very well with the value log$(M_{d}/M_{*})=-2.83\pm0.08$ determined by \cite{Ski11} for all the spirals 
from the KINGFISH survey (see their Table 2). Our value (and the decreasing trend) is also consistent with the value of $-3.03$ found by \cite{Cal17} (see their Table 2) for the KINGFISH spirals,
and the value $~2.0\times10^{-3}$ derived by \cite{Dun11} for low redshift galaxies from the H-ATLAS survey.\\
\cite{Ski11} calculated the dust mass using the dust temperature derived from modified blackbody fits to the far-infrared spectral 
energy distribution of each galaxy, a completely different method to that used in this study. Even so, our results are consistent with those obtained in that study 
for the mean observed dust-to-stellar mass ratio. The corrected mean value, $(0.59\pm0.06)\times10^{-3}$  ($-3.23$ in log scale), is
significantly lower than the measured one, and the one found by \cite{Ski11}.

Another scaling relation that we show in this study is the stellar mass - size relation, in Fig.~\ref{fig:Mstar_Rdisk}. We recover the expected trend - the decimal logarithm of the
stellar mass increases linearly with disk scalelength, e.g. more massive galaxies have more extended stellar disks. We do not correct the stellar masses for dust effects, only the disk 
scalelengths. The corrected relation shows a more accentuated increase of the stellar mass with disk size. This could be caused by the fact that $M_{*}$ values were not corrected to the
intrinsic ones. However, considering that we used stellar mass values from different studies determined using various methods, it was not straightforward to apply corrections because 
for certain wavelengths / emission lines these are simply not available (see Pastrav et al. 2013a, Pastrav et al 2013b). But we emphasize that the respective luminosities or 
nebular / emission line fluxes are affected by dust, to a larger or lesser extent. Nevertheless, obtaining an accurate mass - size relation is important for galaxy evolution studies.

\begin{figure*}
\begin{center}
 \includegraphics[scale=0.47]{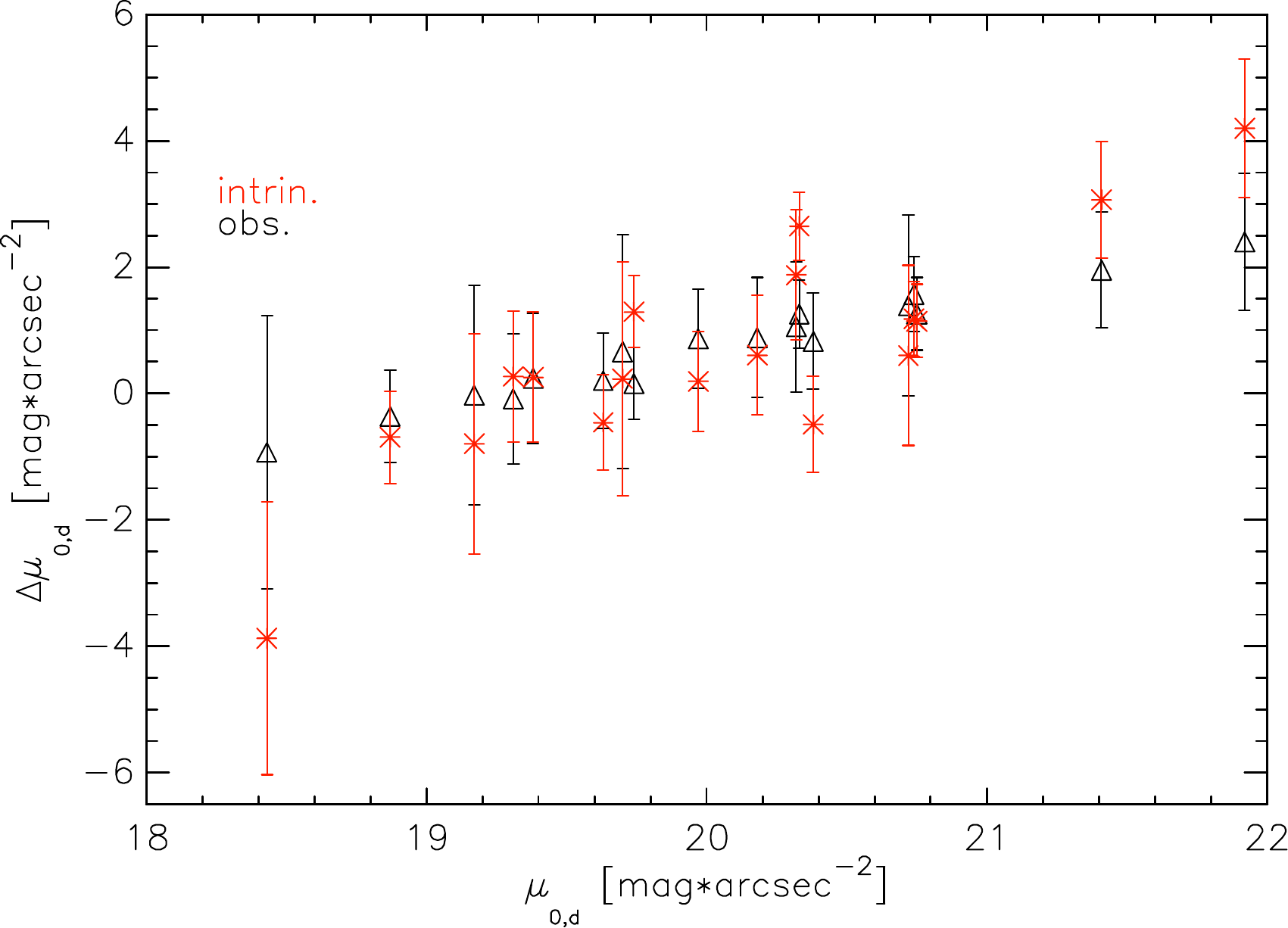}
 \hspace{-0.0cm}
 \includegraphics[scale=0.47]{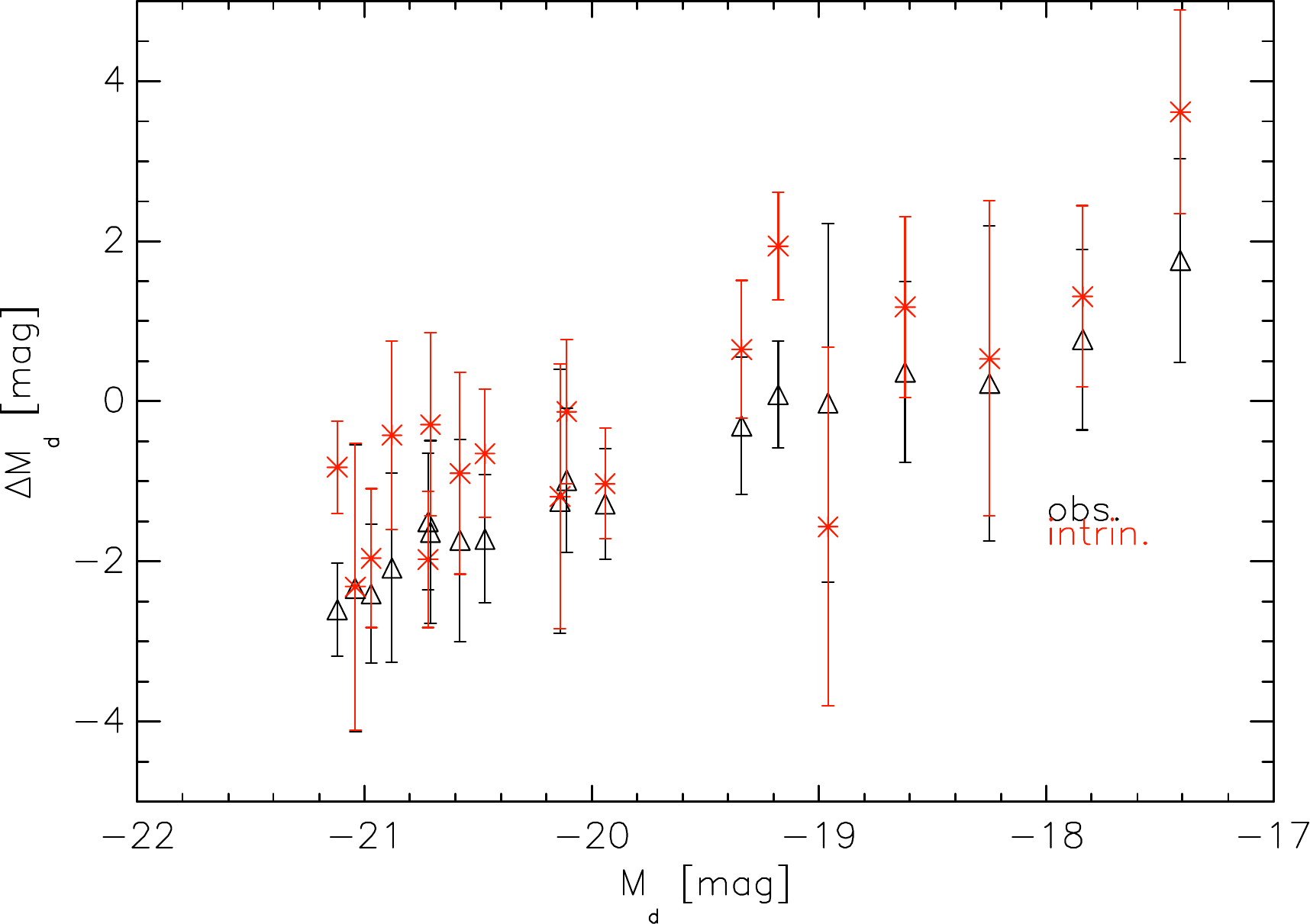}
 \caption{\label{fig:residuals_mu0_Mdisk} Left panel: Residuals, between observed central disk surface brightness values and the linear regression fit to the observed central disk surface 
 brightness - disk scalelength relation (see Fig.~\ref{fig:mu0_Mdisk}), in black color; in red color, the same but for the intrinsic case. Right panel: The same, but for the disk absolute magnitude - 
 disk scalelength relation. The symbols are the same as in previous plots.}
 \end{center}
\end{figure*}

\section{Discussion}\label{sec:discussion}

One important feature of any scaling relation is its scatter, which a succesful model of galaxy evolution should be able to reproduce, and the zero point and the slope
of the relation. The cause of the low scatter in scaling relations such as Tully-Fisher, Faber Jackson and others is still not fully established. A contribution to this scatter
may come from innacurately derived parameters involved in these relations (due to fitting limitations of each galaxy) or not taking into account biases introduced by dust.
In this respect, we analyse the size-luminosity relation to see if applying dust and inclination corrections produce either a reduction or an increase in the scatter of the intrinsic relation 
compared to the observed one. Specifically, we analyse the scatter of the surface brightness - scalelength or equivalently, of the absolute magnitude - scalelength relations. 
For this pupose, we calculate the residuals: the observed (intrinsic) values minus the linear regression fit to the observed (intrinsic) values. The corresponding values
are presented in Fig.~\ref{fig:residuals_mu0_Mdisk}, with the corresponding uncertainties overplotted. \\
As one can see from both plots, there is a very mild dependence of the residuals on the central surface brightness (absolute magnitude). However, taking into account
the uncertainties and leaving out the two data points which are far out from the fits also in Figs.~\ref{fig:mu0_Mdisk}-\ref{fig:Mstar_Rdisk}, one cannot observe a clear decrease or 
increase in the scattering of these relations when going from the observed to the intrinsic ones.
At longer wavelengths we expect to have even flatter trends in the variation of the residuals with central surface brightess (absolute magnitude), as the combined effects of dust 
and inclination are less pronounced.\\
We also test the correlation for $\mu_{0,d}-R_{s,d}$ and $M_{d}-R_{s,d}$ relations by calculating the Pearson correlation coefficents. Thus, for the surface brightness - size 
relation we get $r=0.46$ for the measured values and $r=0.42$ for the intrinsic ones, while for the magnitude - size relation we obtain Pearson coefficients of $-0.76$ and $-0.40$.
As the coefficients for the measured relations are closer to 1 or -1 than the ones for the corrected relations, it appears that at this level, eliminating the bias introduced by 
projection, dust and decomposition effects produces less correlated size-luminosity relations and the scatter is increased. Hovever, we need to take into consideration that 
our sample is quite small and limited to nearby (low redshift) galaxies.\\

If we were to analyse a similar sample of spiral galaxies at higher redshifts but at same wavelength, in order to assess how these relations would change, we would notice
that the disk scalelengths would be smaller considering an inside-out scenario of galaxy evolution. Correspondingly, the disk central surface brightness and magnitude will be higher as 
we go towards higher redshifts. This has been shown by \cite{Bar05}, by analysing the surface brightness and surface mass density evolution of a sample of disk galaxies from the GEMS 
(Galaxy Evolution from Morphology and SEDs) survey, in V band and at redshifts up to $z=1.1$. \cite{Bar05} find a ~1 mag increase in magnitudes and central surface brightness by $z=1$. 


\section{Summary and conclusions}\label{sec:conclusions}

We presented here the results of a study of the combined effects of dust, inclination and decomposition effects on some of the scaling relations of nearby spiral galaxies.
We have done a detailed analysis of a sample of 18 typical unbarred spiral galaxies taken from the KINGFISH survey, in B band, representative for the nearby universe galaxies. 
A careful surface photometry and sky determination and subtraction was performed to derive the integrated fluxes of each galaxy and its main components - disks and bulges.
We derived the measured (observed) photometric and structural parameters of disks and bulges by doing a 2-component bulge-disk decomposition of galaxies, using GALFIT data analysis
algorithm. Prior to fitting, a customized mask was applied to each galaxy image to take out all the bad pixels corresponding to nearby stars/galaxies, image artifacts, negative pixels
and noise. Together with the intergrated fluxes and the structural parameters, we calculated the central surface brightness and absolute magnitude for disks and bulges of each galaxy,
and the bulge-to-disk ratios.\\

Using \cite{Gro13} empirical relation (Eq.~\ref{eq:Grootes}), we have derived the central face-on dust opacities in B band ($\tau_{B}^{f}$), and subsequently the dust mass of each 
galaxy, using Eq.~\ref{eq:Mdust} - obtained considering the dust geometry of \cite{Pop11} model. Using the numerical corrections for projection, dust and decomposition effects 
determined in \cite{Pas13a} and \cite{Pas13b} we corrected all the necessary photometric and strucutural parameters to obtain their intrinsic values. We then presented disk scaling
relations, such as central surface brightness - scalelength, magnitude - scalelength, dust vs stellar mass, dust-to-stellar ratios vs. stellar mass/ stellar mass surface density, or 
stellar mass as a function of disk scalelength. Both the observed (measured) and the intrinsic (corrected) relations were presented, corrected for all the aforementioned effects, in order to 
better illustrate the differences in the overall trend, in their slopes and zero points. By analysing these relations, our main conclusions are:
\begin{itemize}
 \item for the size-luminosity type of relations (Fig.~\ref{fig:mu0_Mdisk}), the decrease in central surface brightness is important as a result of 
 applying the corrections - up to \textit{~2 mag.} or more, in some cases; the corresponding changes in the values of $M_{d}$ are also quite large, but not as much as 
for $\mu_{0,d}$
 \item the slope of the $\mu_{0,d} - R_{s,d}$ relation changes from $0.32$ (observed) to $0.45$ (intrinsic), while for the $M_{d} - R_{s,d}$ relation the slope 
 decreases from $-0.46$ to $-0.32$, when corrections are applied; the slopes for the observed relations are similar with those found in other studies of spiral galaxies
 \item  the slope of $M_{dust}-M_{*}$ is close to unity for the observed relation - $1.04\pm0.02$, with a small decrease for the corrected one ($0.98\pm0.14$)
 \item the mean value for the central face-on dust opacity that we find for our small sample - $\tau_{B}^{f}=3.71\pm0.43$, is consistent with other studies of dust 
 attenuation in spiral galaxies, done on much larger samples, such as the one of \cite{Dri07}
 \item we recover the expected trends in the dust-to-stellar mass ratio as a function of stellar mass / stellar surface density (Fig.~\ref{fig:DS_Mstar_ustar})
 only after applying the necessary corrections, which comes to underline the importance of having unbiased, dust free scaling relations
 \item the mean value of the observed dust-to-stellar mass ratio of our sample, $(1.39\pm0.04)\times10^{-3}$ ($-2.85$ in log scale) is in very good agreement with 
 the mean value log$(M_{d}/M_{*})=-2.83\pm0.08$ found by \cite{Ski11} for all the spirals from the same KINGFISH survey, using a completely different method; 
 the intrinsic value, $(0.59\pm0.06)\times10^{-3}$ ($-3.23$ in log scale), is lower than the measured one, and the one found by \cite{Ski11}
 \item the intrinsic size-luminosity relations show a larger scatter than the observed ones, while the correlations are not so strong
\end{itemize}
We have chosen to do this study in B band as our method is useful when optical data is available, but also because we can see more clearly the changes in the scaling relations at 
shorter wavelengths, when dust effects are stronger. Thus, the method presented here, to obtain intrinsic scaling relations, is tailored for cases where
optical data is available. In this paper we have concentrated on disk scaling relations and spiral galaxies. In a future paper, we will show corresponding
results for bulges and elliptical galaxies.

\section*{Acknowledgements}
We would like to thank the referee, Stephen Serjeant, for a careful reading of the manuscript and for the useful suggestions 
which improved the quality and clarity of this paper. This research made use of the NASA/IPAC Extragalactic Database (NED), which
is operated by the Jet Propulsion Laboratory, California Institute of Technology, under contract with the National Aeronautics and Space Administration.\\
The author would like to acknowledge financial support from the project Nucleu - LAPLAS VI.

\appendix
\section{Photometry of the sample}
We show here the figures analogous to Figs.~\ref{fig:NGC3031}\&\ref{fig:NGC4826}, corresponding to the other 16 galaxies from our sample.
Some considerations about the galaxies and the fits were added in the corresponding figure captions. While for some galaxies the fits to the
observed surface brightness profiles are not perfect, one needs to take into account the particular features each galaxy image presented, 
which could not always be accurately modeled with a combination of an exponential and a S\'{e}rsic functions only. 

\begin{figure*}
 \begin{center}
  \includegraphics[angle=90,origin=c]{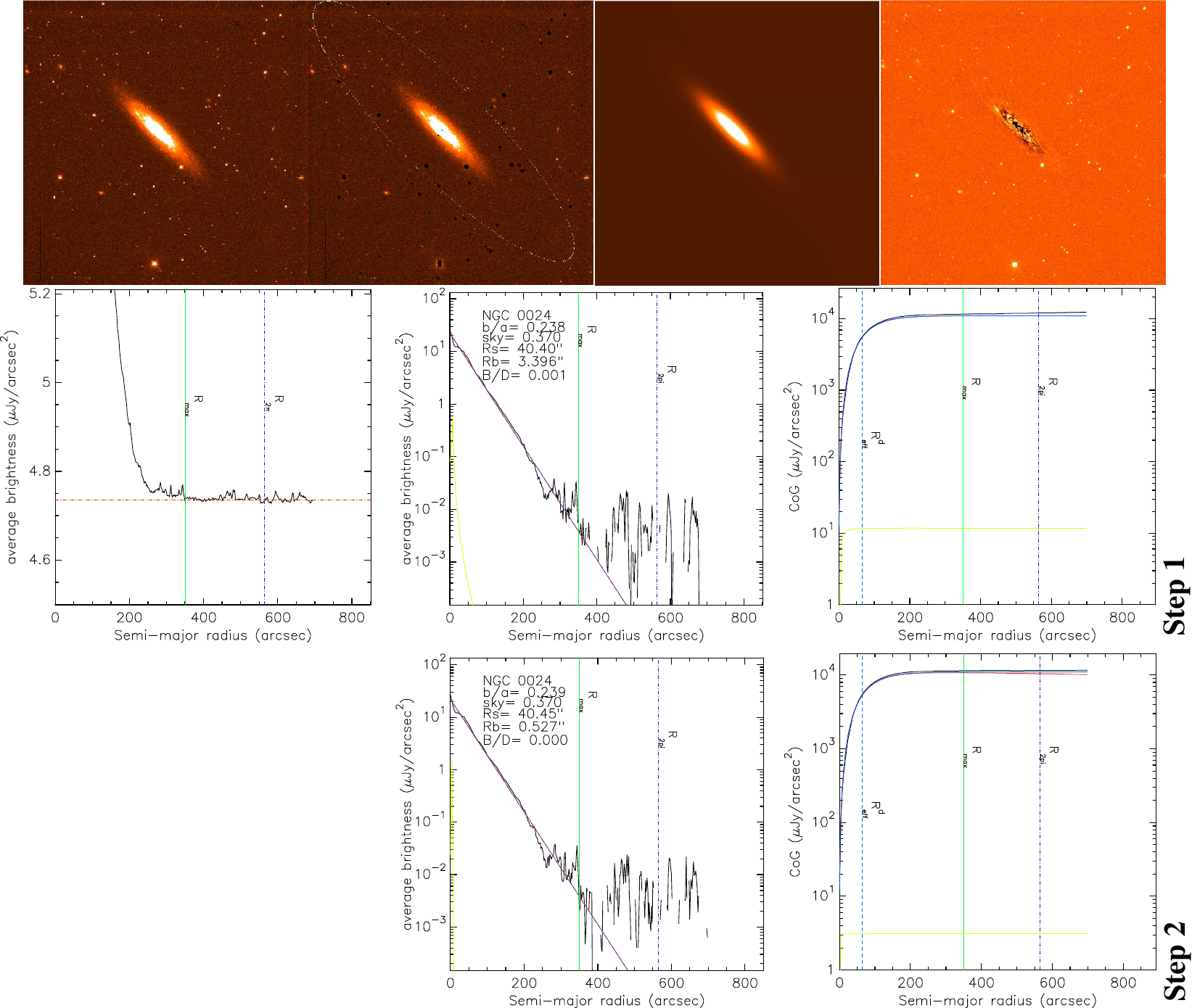}
  \caption{\label{fig:NGC0024} \textbf{NGC0024} As for \ref{fig:NGC3031}. The angular size of the observed image on the sky is 
   $14.82^{\prime}\times14.82^{\prime}$. This is basically a bulgeless galaxy. In the top row - 2nd panel from the left, the dashed
   white ellipse denotes $R_{2\pi}$, the major axis radius out to which data is available over the full azimuthal range. Step 3 
   fit was not neccesary for this galaxy.}
 \end{center}
\end{figure*}
\newpage
\begin{figure*}
 \begin{center}
  \includegraphics[angle=90,origin=c]{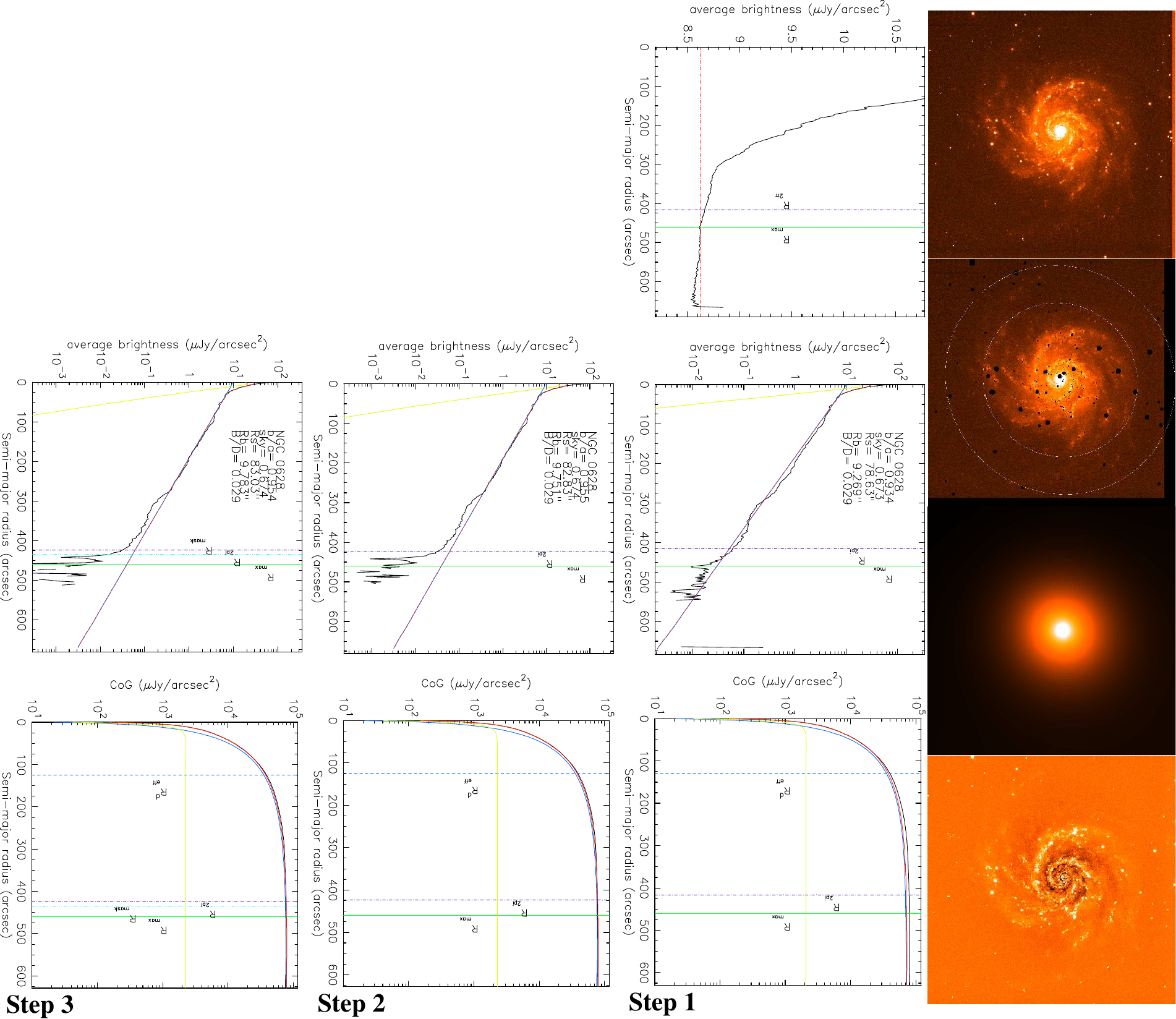}
  \caption{\label{fig:NGC0628} \textbf{NGC0628} As for Fig.~\ref{fig:NGC4826}. The angular size of the observed image on the sky is $14.82^{\prime}\times14.82^{\prime}$.}
 \end{center}
\end{figure*}
\newpage
\begin{figure*}
 \begin{center}
  \includegraphics[angle=90,origin=c]{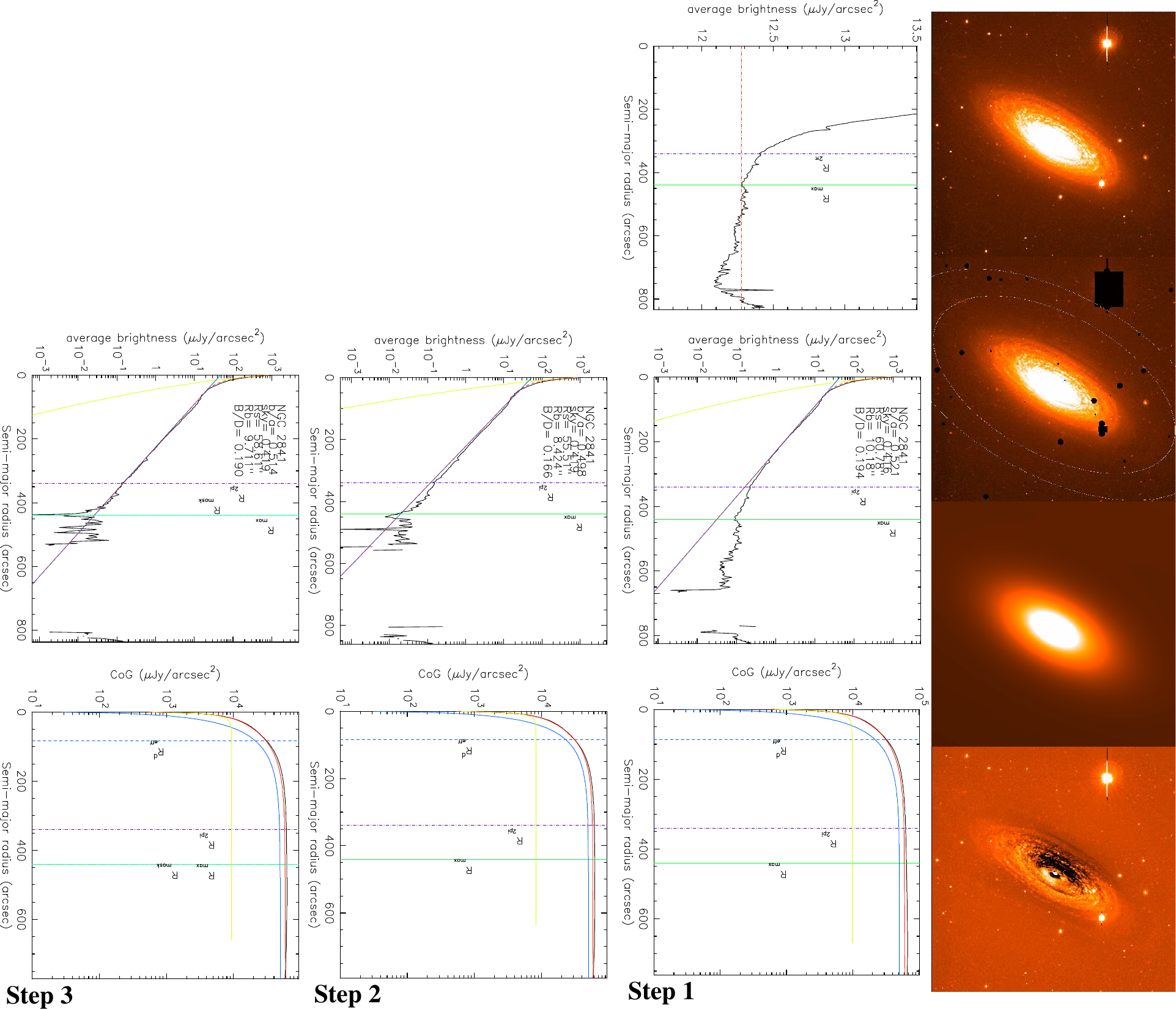}
  \caption{\label{fig:NGC2841} \textbf{NGC2841} As for Fig.~\ref{fig:NGC4826}. In the top row - 2nd panel from the left, the two dashed white ellipses denote
   $R_{2\pi}$ (inner ellipse, the major axis radius out to which data is available over the full azimuthal range and $R_{max}$ (outer ellipse, 
   the semi-major radius out to which emission from the galaxy could be detected). The observed image is potentially not large enough for a determination 
   with greater precision of the background. The angular size of the observed image on the sky is $10.38^{\prime}\times10.36^{\prime}$.}
 \end{center}
\end{figure*}
\newpage
\begin{figure*}
 \begin{center}
  \includegraphics[angle=90,origin=c]{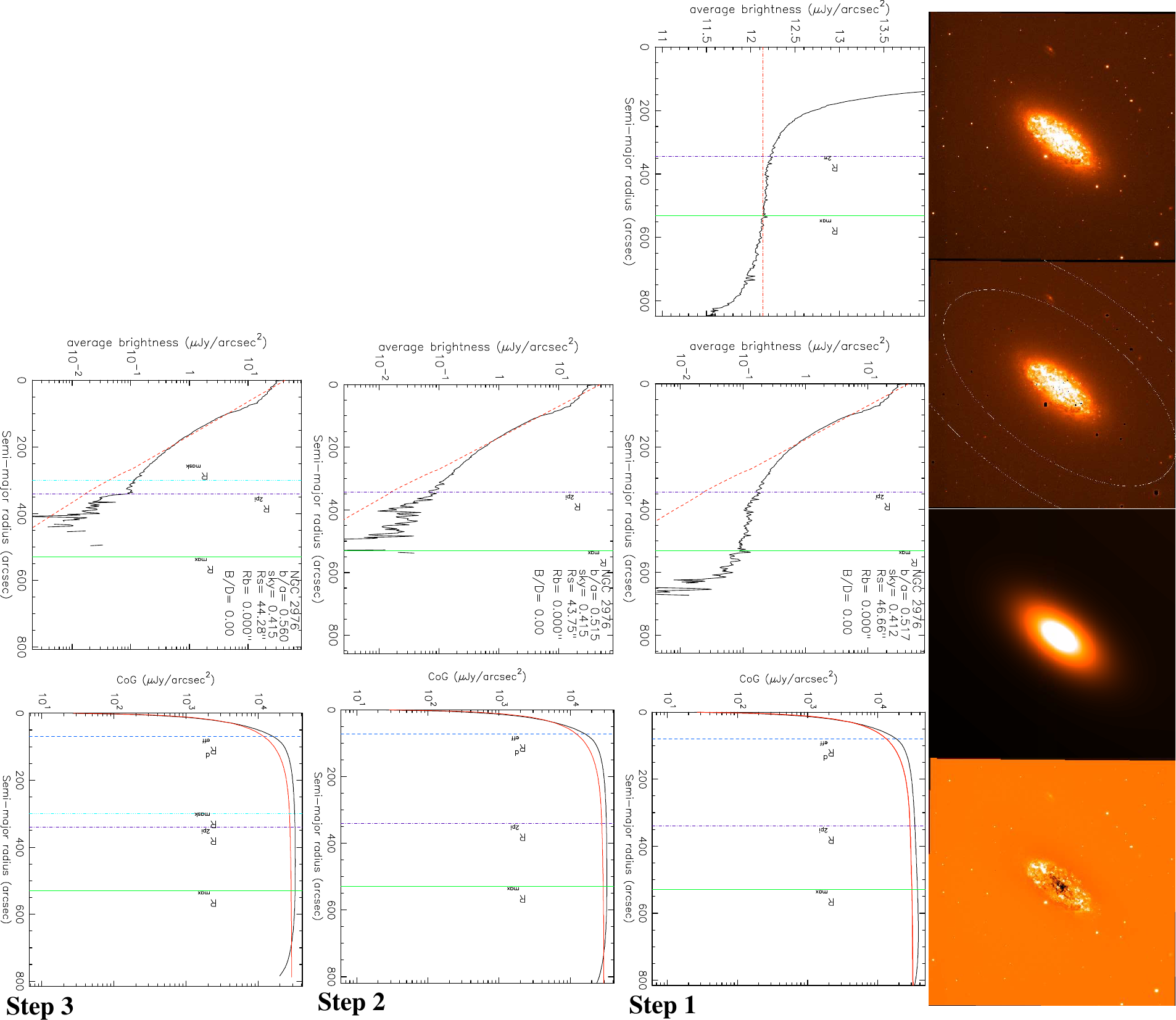}
  \caption{\label{fig:NGC2976} \textbf{NGC2976} As for Fig.~\ref{fig:NGC4826}. This is essentially a bulgeless galaxy. Fitting with two components did not result a clear
  bulge. The angular size of the observed image on the sky is $10.43^{\prime}\times10.45^{\prime}$.}
 \end{center}
\end{figure*}
\newpage
\begin{figure*}
 \begin{center}
  \includegraphics[angle=90,origin=c]{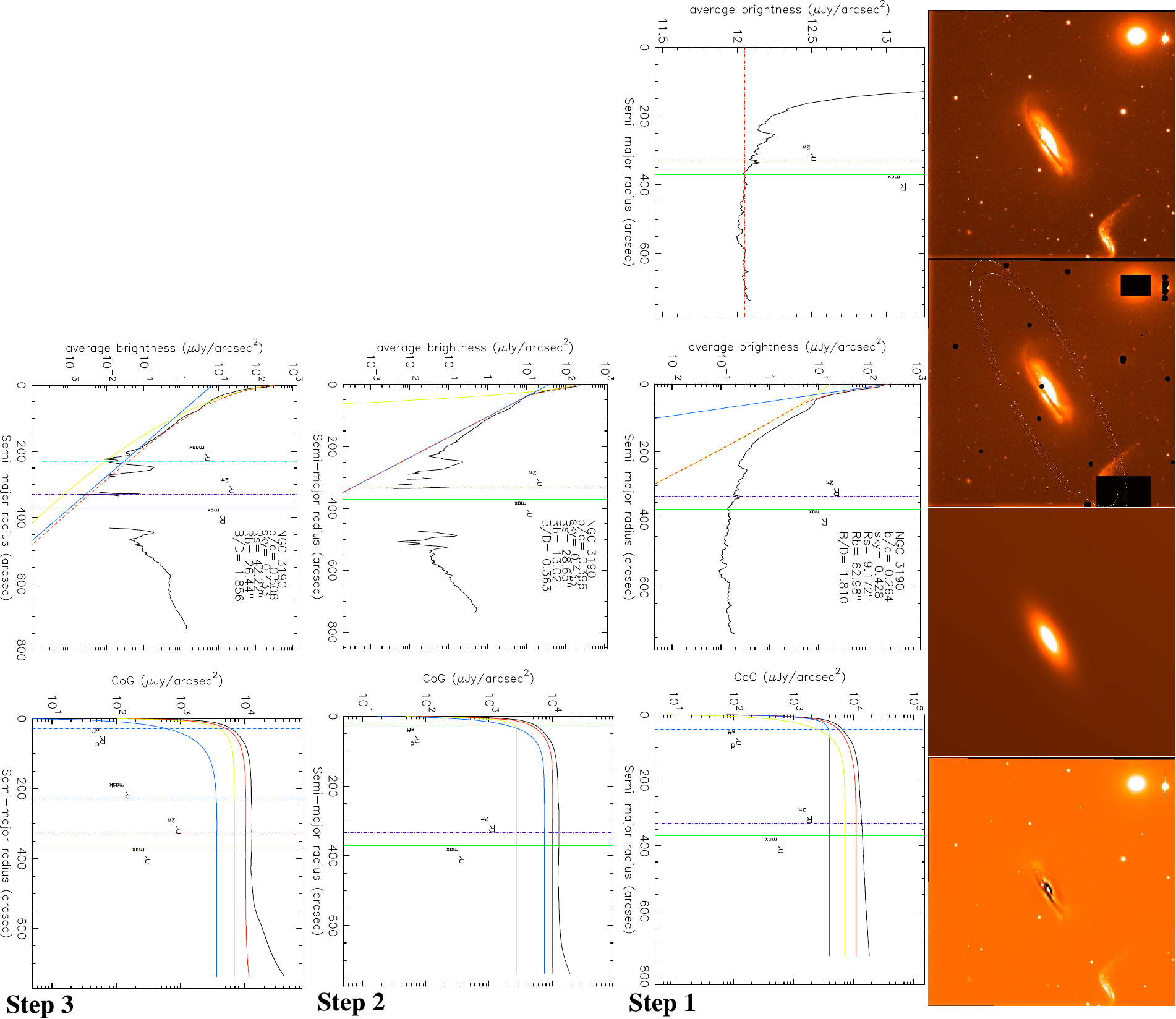}
  \caption{\label{fig:NGC3190} \textbf{NGC3190} As for Fig.~\ref{fig:NGC4826}. This is a peculiar type of galaxy. The dust lane that surrounds the stellar disk
  creates difficulties in recovering an accurate surface brightess profile with a two-component fit. The angular size of the observed image on the sky is $10.43^{\prime}\times10.43^{\prime}$.}
 \end{center}
\end{figure*}
\newpage
\begin{figure*}
 \begin{center}
  \includegraphics[angle=90,origin=c]{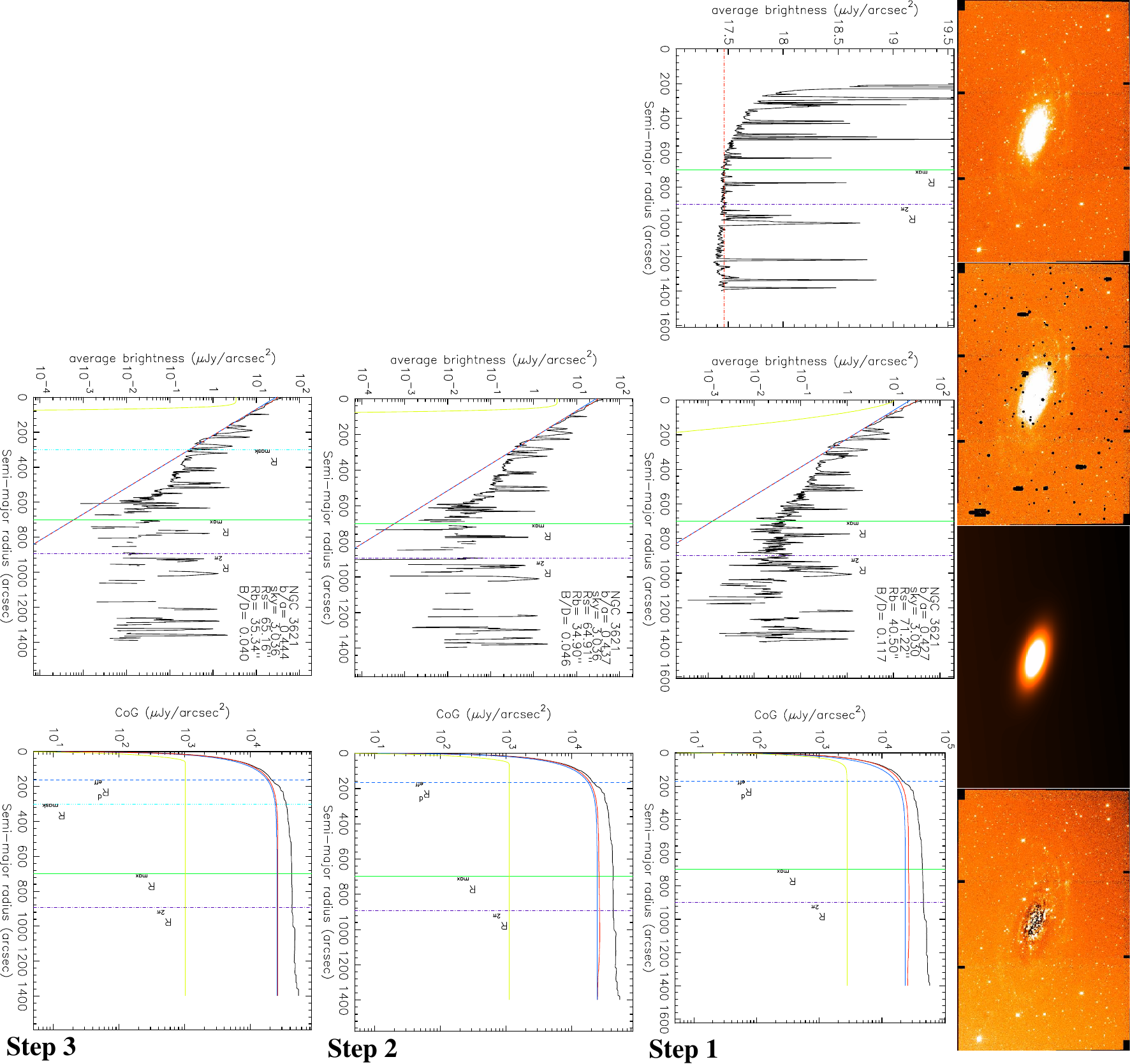}
  \caption{\label{fig:NGC3621} \textbf{NGC3621} As for Fig.~\ref{fig:NGC4826}. The observed image for this galaxy was a large combined mozaic of multiple images,
  with a significant degree of noise. However, for the relevant galactocentric radii (up to 300 arcsecs., beyond which there is esentially
  noise and almost no galaxy emission), we managed to fit the surface brightness profile. The angular size of the observed image on the 
  sky is $29.08^{\prime}\times19.12^{\prime}$.}
 \end{center}
\end{figure*}
\newpage
\begin{figure*}
 \begin{center}
  \includegraphics[angle=90,origin=c]{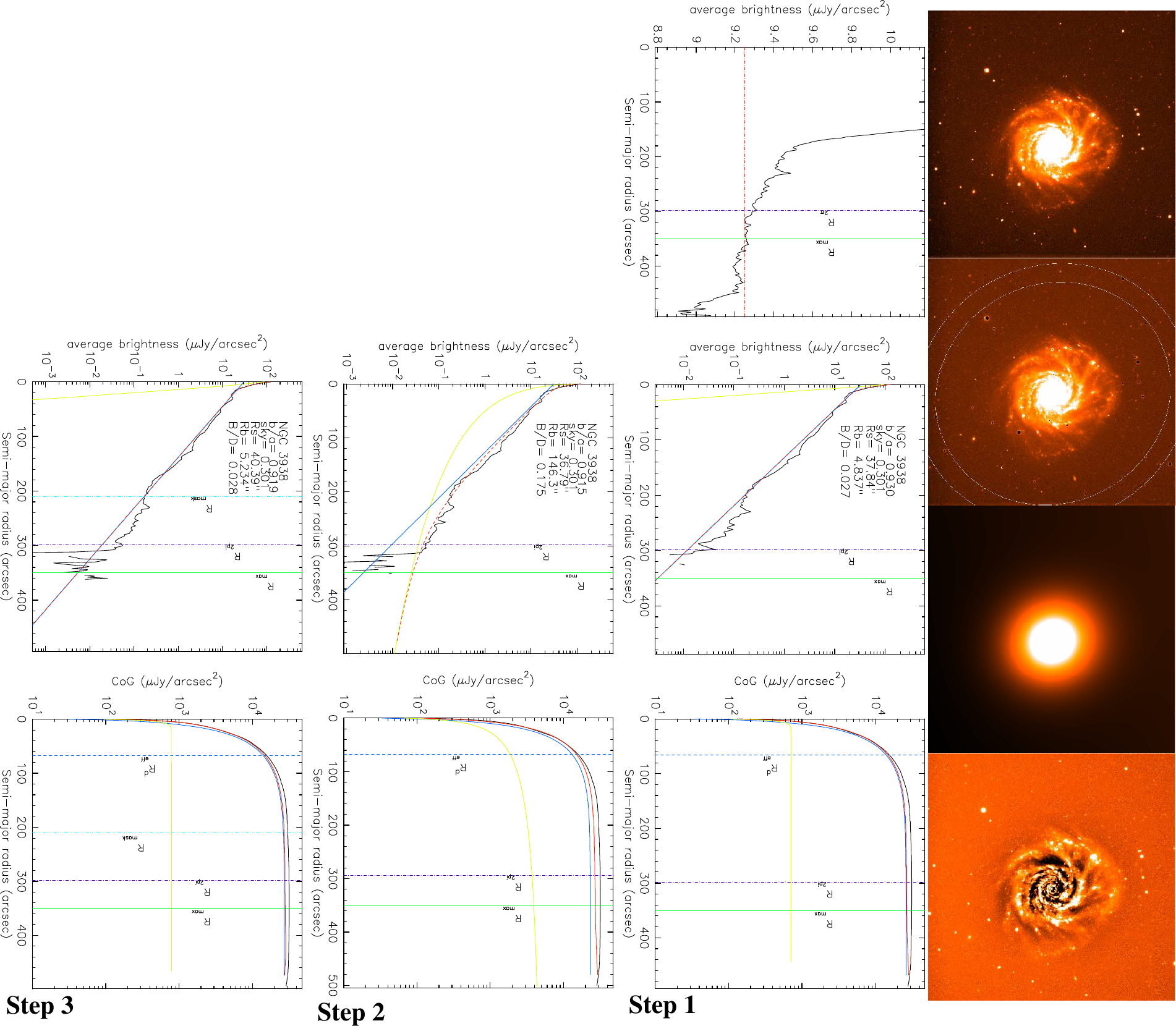}
  \caption{\label{fig:NGC3938} \textbf{NGC3938} As for Fig.~\ref{fig:NGC4826}. The angular size of the observed image on the sky is $9.88^{\prime}\times9.88^{\prime}$.}
 \end{center}
\end{figure*}
\newpage
\begin{figure*}
 \begin{center}
  \includegraphics[angle=90,origin=c]{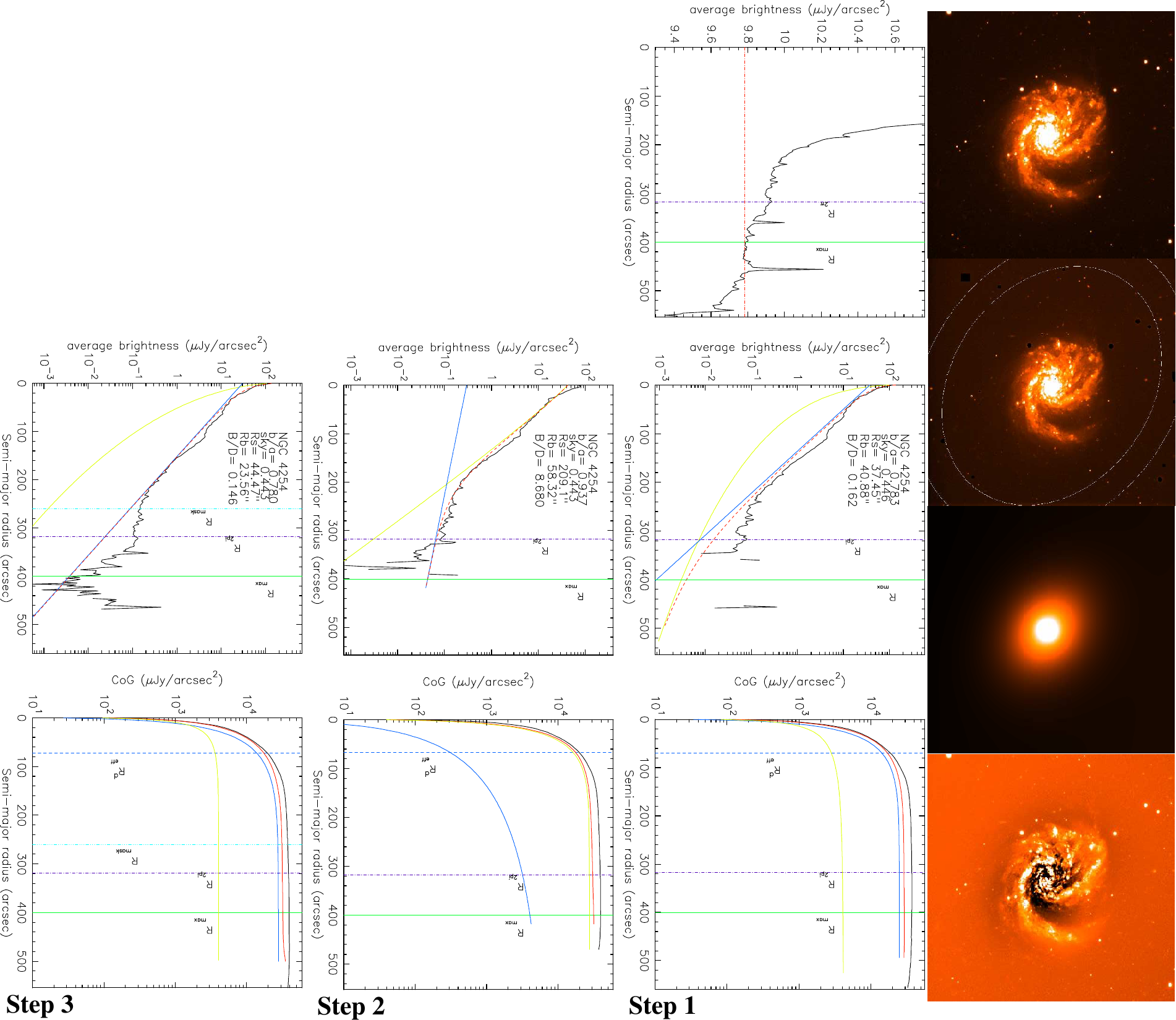}
  \caption{\label{fig:NGC4254} \textbf{NGC4254} As for Fig.~\ref{fig:NGC4826}. The angular size of the observed image on the sky is $10.36^{\prime}\times10.37^{\prime}$.}
 \end{center}
\end{figure*}
\newpage
\begin{figure*}
 \begin{center}
  \includegraphics[angle=90,origin=c]{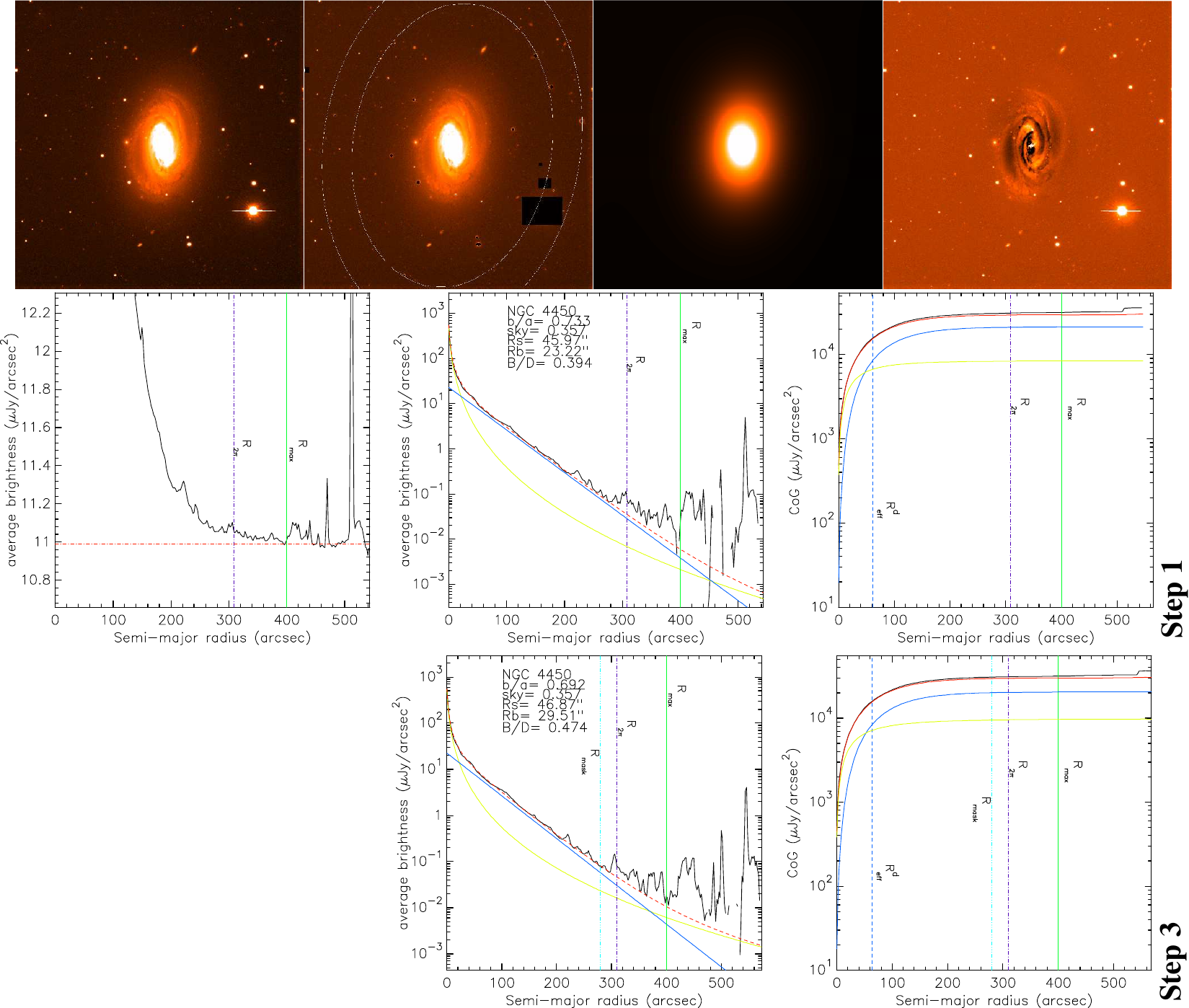}
  \caption{\label{fig:NGC4450} \textbf{NGC4450} As for Fig.~\ref{fig:NGC4826}. Step 2 fits are not presented, as the fit was not meaningful. The angular size of the observed 
  image on the sky is $10.36^{\prime}\times10.36^{\prime}$.}
 \end{center}
\end{figure*}
\newpage
\begin{figure*}
 \begin{center}
  \includegraphics[angle=90,origin=c]{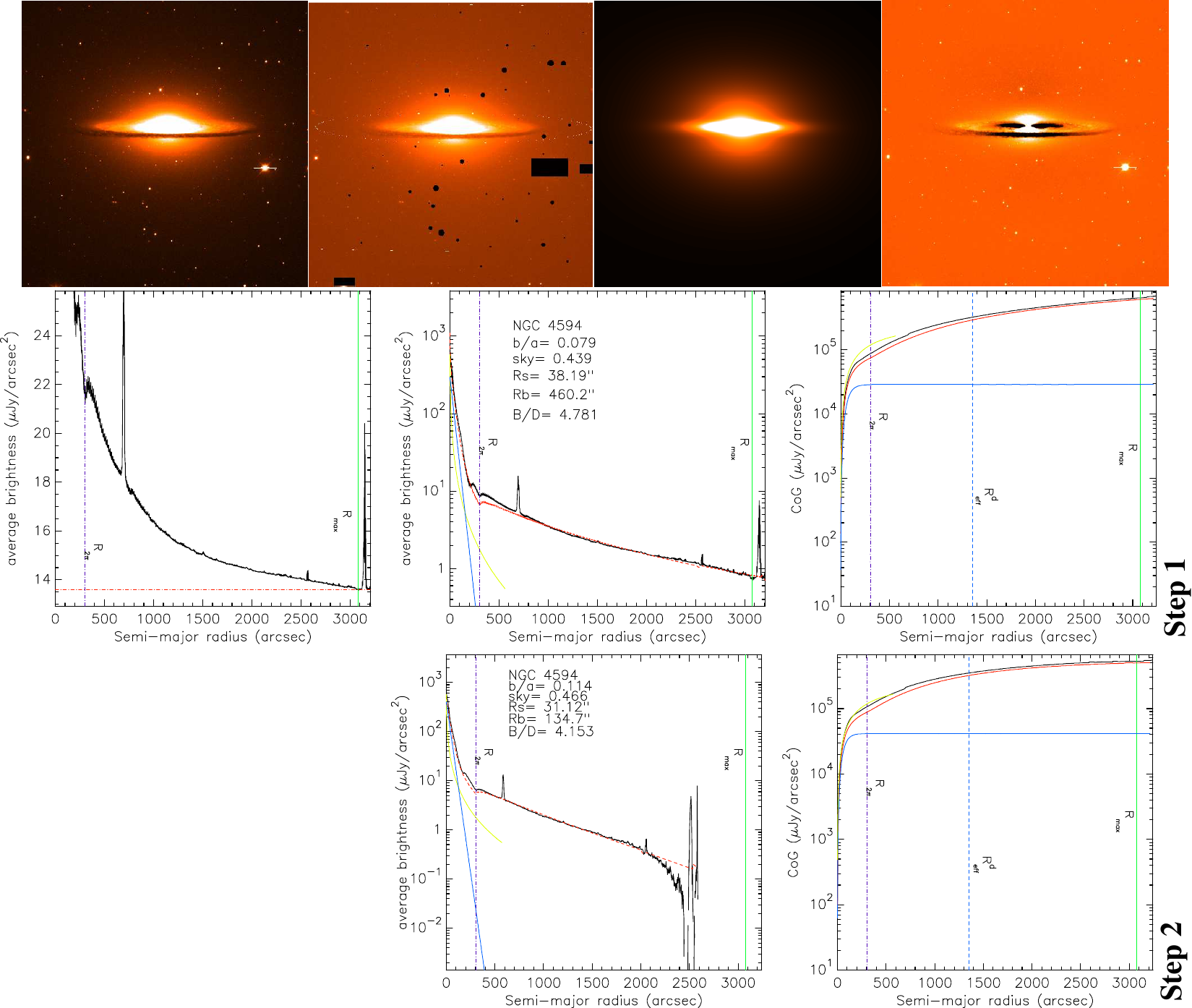}
  \caption{\label{fig:NGC4594} \textbf{NGC4594} As for Fig.~\ref{fig:NGC3031}. This is the so called ``Sombrero Galaxy'', with a huge bulge, divided by the dust and stellar disks,
  and a large value for $B/D$. The fitting procedure was particularly complicated but we believe the results are meaningful. The angular size of the observed image on the sky is
  $10.36^{\prime}\times10.37^{\prime}$.}
 \end{center}
\end{figure*}
\newpage
\begin{figure*}
 \begin{center}
  \includegraphics[angle=90,origin=c]{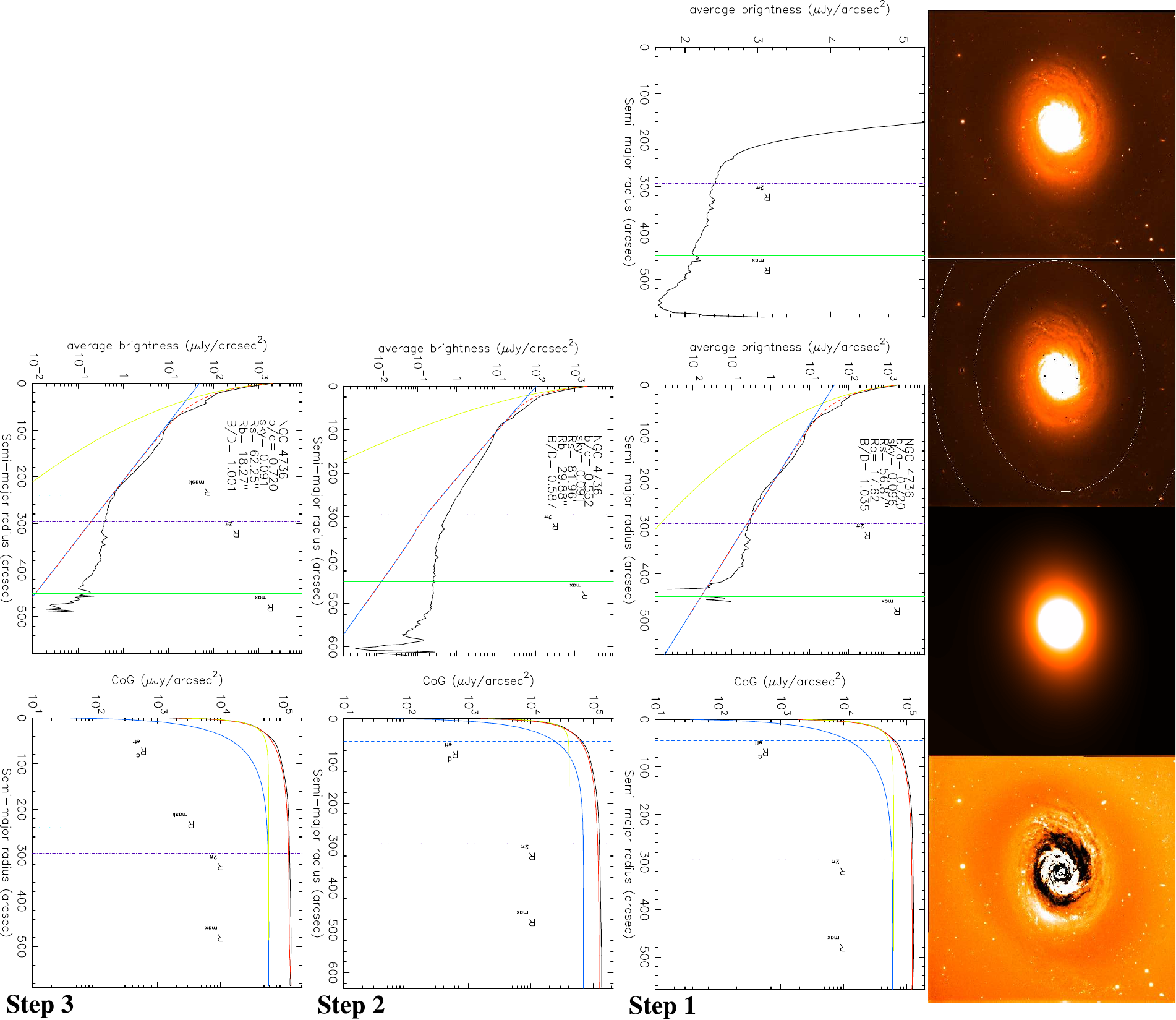}
  \caption{\label{fig:NGC4736} \textbf{NGC4736} As for Fig.~\ref{fig:NGC4826}. The angular size of the observed image on the sky is $10.41^{\prime}\times10.43^{\prime}$.}
 \end{center}
\end{figure*}
\newpage
\begin{figure*}
 \begin{center}
  \includegraphics[angle=90,origin=c]{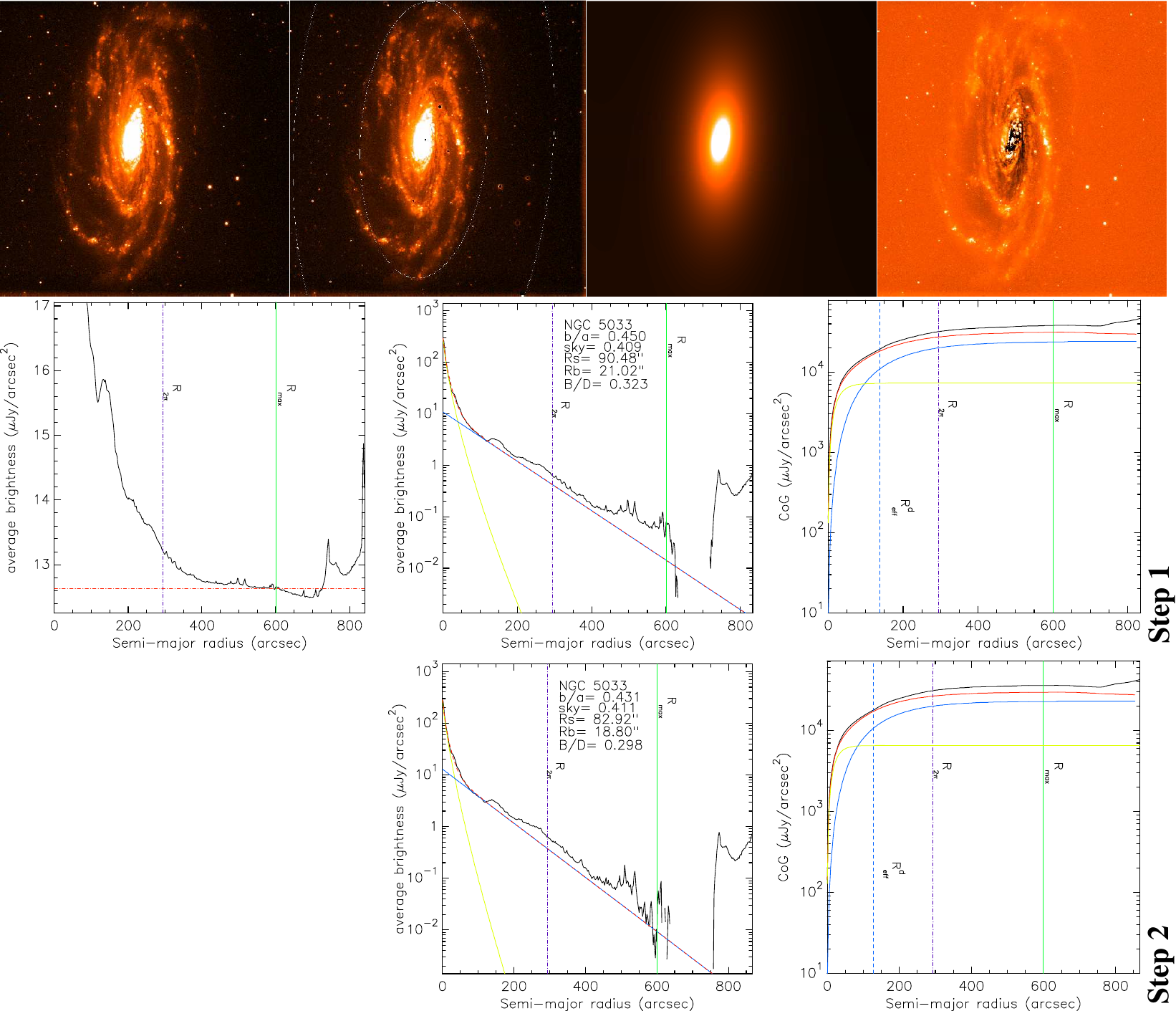}
  \caption{\label{fig:NGC5033} \textbf{NGC5033} As for Fig.~\ref{fig:NGC3031}. The angular size of the observed image on the sky is
  $10.36^{\prime}\times10.37^{\prime}$.}
 \end{center}
\end{figure*}
\newpage
\begin{figure*}
 \begin{center}
  \includegraphics[angle=90,origin=c]{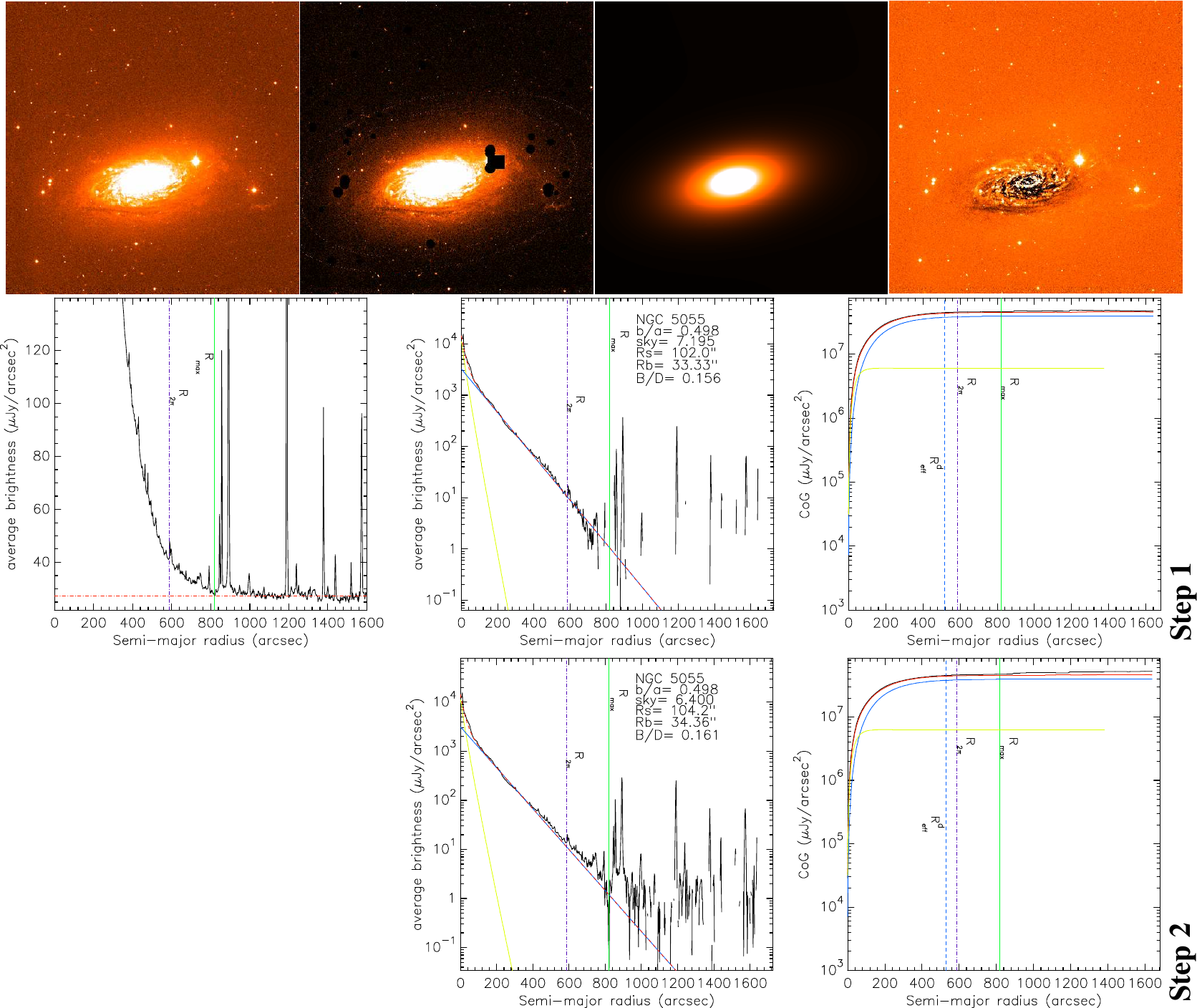}
  \caption{\label{fig:NGC5055} \textbf{NGC5055} As for Fig.~\ref{fig:NGC3031}. The angular size of the observed image on the sky is
  $20.30^{\prime}\times20.30^{\prime}$.}
 \end{center}
\end{figure*}
\newpage
\begin{figure*}
 \begin{center}
  \includegraphics[angle=90,origin=c]{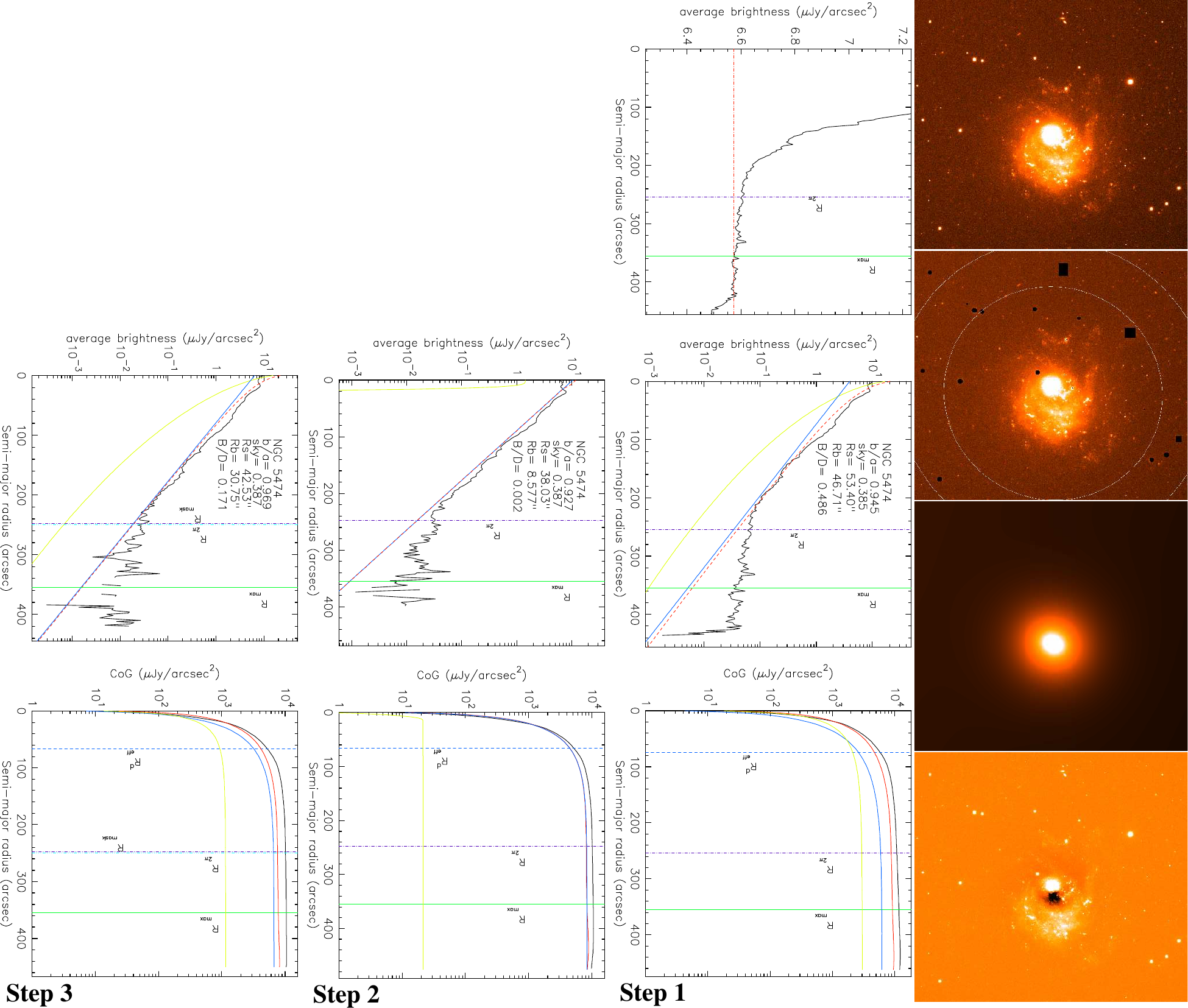}
  \caption{\label{fig:NGC5474} \textbf{NGC5474} As for Fig.~\ref{fig:NGC4826}. This is an asymmetric galaxy, probably disturbed because of a merger.
  The fitting procedure was a bit more complex. The angular size of the observed image on the sky is $18.50^{\prime}\times20.31^{\prime}$.}
 \end{center}
\end{figure*}
\newpage
\begin{figure*}
 \begin{center}
  \includegraphics[angle=90,origin=c]{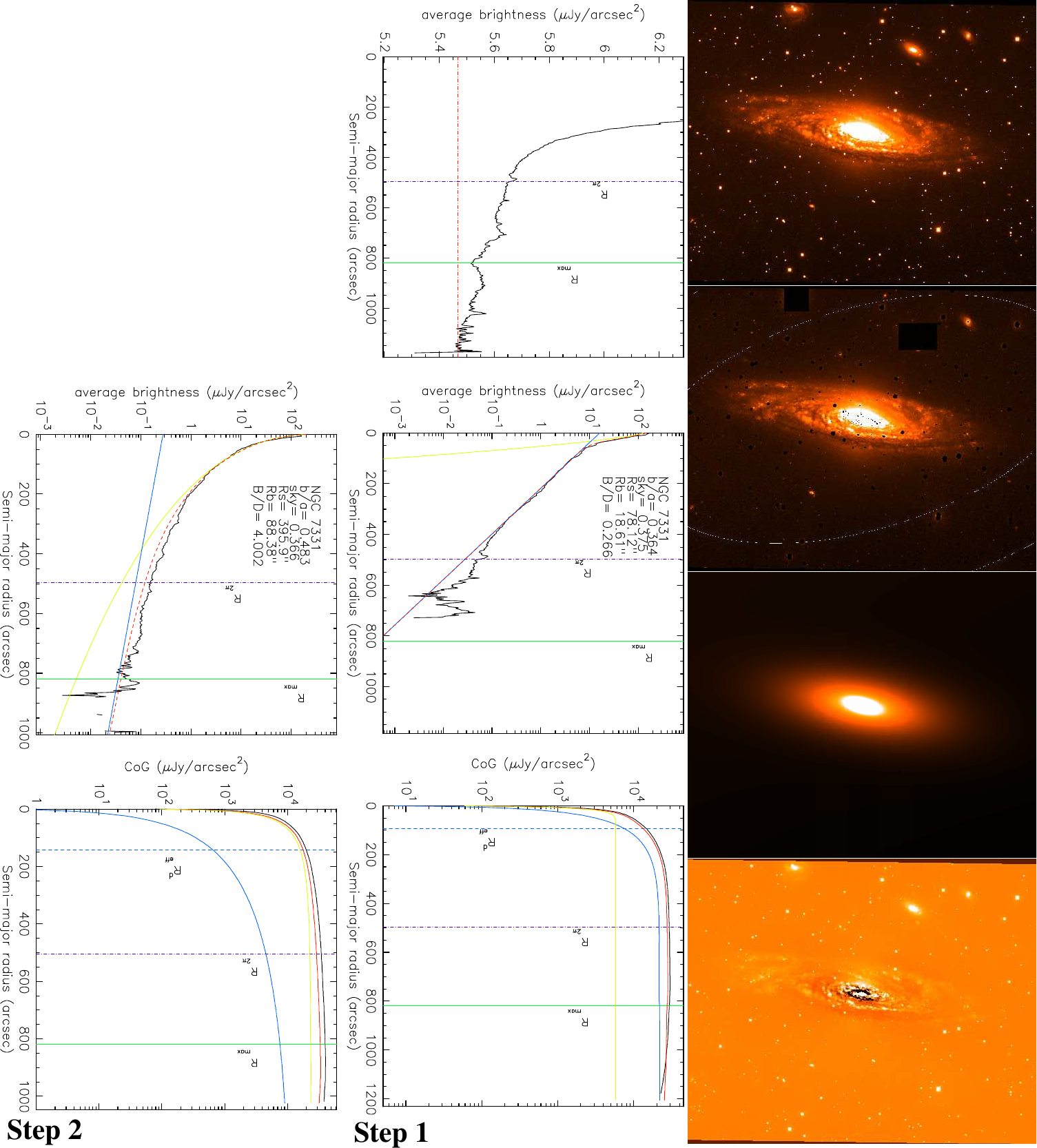}
  \caption{\label{fig:NGC7331} \textbf{NGC7331} As for Fig.~\ref{fig:NGC3031}. The angular size of the observed image on the sky is
  $10.59^{\prime}\times16.40^{\prime}$.}
 \end{center}
\end{figure*}
\newpage
\begin{figure*}
 \begin{center}
  \includegraphics[angle=90,origin=c]{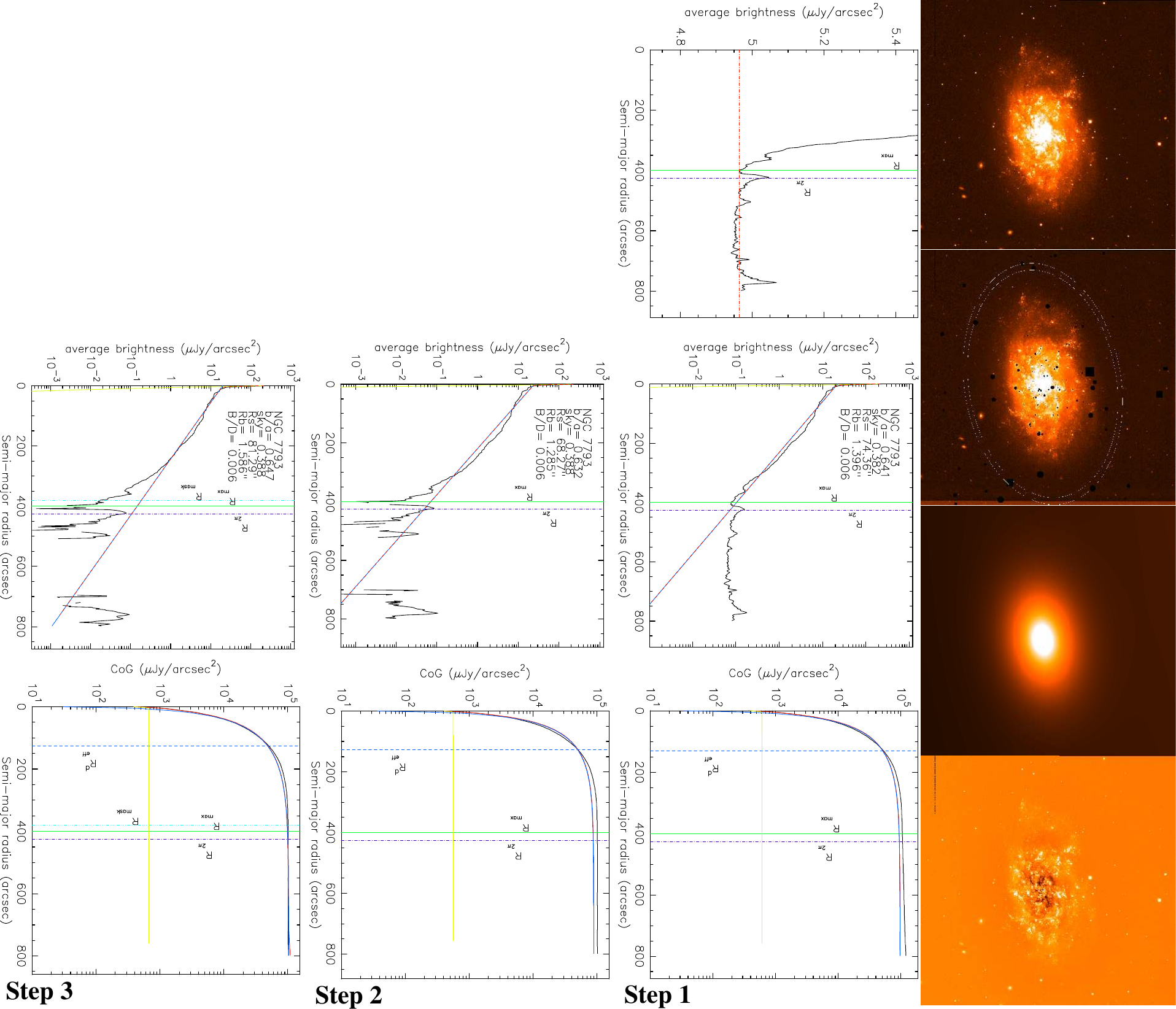}
  \caption{\label{fig:NGC7793} \textbf{NGC7793} As for Fig.~\ref{fig:NGC4826}. The angular size of the observed image on the sky is
  $14.83^{\prime}\times14.83^{\prime}$.}
 \end{center}
 \end{figure*}
 
\bsp	
\label{lastpage}

\end{document}